\documentclass[a4paper,fleqn,usenatbib]{mnras}

\usepackage{newtxtext,newtxmath}

\usepackage[T1]{fontenc}
\usepackage{ae,aecompl}

\usepackage{graphicx,epsfig,times,color,amsmath,amssymb,pifont}

\usepackage{subfig}

\usepackage{enumerate}

\usepackage{url}

\usepackage{hyperref}

\def\Msun{\mbox{~M$_\odot$}}

\def\kms{\mbox{~km~s$^{-1}$}}

\def\kpc{\mbox{~kpc}}
\def\Mpc{\mbox{~Mpc}}

\def\vcirc{V_{circ}}
\def\vmax{V_{\rm max}}
\def\vlos{V_{\rm los}}
\def\vout{V_{\rm out}}
\def\vrot{V_{\rm rot}}

\def\Vcirc{V_{\rm circ}}
\def\vcirc{V_{\rm circ}}
\def\Vmax{V_{\rm max}}
\def\Vlos{V_{\rm los}}
\def\Vout{V_{\rm out}}
\def\Vrot{V_{\rm rot}}

\def\Rvir{R_{\rm vir}}
\def\Mvir{M_{\rm vir}}

\def\Mbar{M_{\rm bar}}

\def\Mstar{M_{\rm star}}
\def\Mhalo{M_{\rm halo}}

\def\apj{\rm Astrophys. J.}
\def\aap{\rm Astron. \&Astrophys. }
\def\mnras{\rm Mon.\,Not.\,RAS. }

\def\Msunh{h^{-1}M_\odot}

\def\mathnew{\mathsurround=0pt}
\def\simov#1#2{\lower .5pt\vbox{\baselineskip0pt
    \lineskip-.5pt\ialign{$\mathnew#1\hfil##\hfil$\crcr#2\crcr\sim\crcr}}}

\def\lesssim{\mathrel{\mathpalette\simov <}}
\def\apj{\rm ApJ}
\def\apjl{\rm ApJL}
\def\apjs{\rm ApJS}
\def\aj{\rm AJ}
\def\mnras{\rm MNRAS}
\def\nat{\rm Nature}

\def\aap{\rm A\&A}

\newcommand{\HI}{\hbox{{\sc H}\hspace{0.7pt}{\sc i}} }

\title[Abundance of the faintest galaxies in CDM]{Another baryon miracle? Testing solutions to the ``missing dwarfs'' problem}

\author[S. Trujillo-Gomez et al.]
{Sebastian Trujillo-Gomez$^{1,2}$\thanks{E-mail: strujill@gmail.com}, 
Aurel Schneider$^3$,
Emmanouil Papastergis$^4$, 
\newauthor Darren S. Reed$^{1,2,5}$,
George Lake$^{1,2}$ \\  \
$^1$Center for Theoretical Astrophysics and Cosmology, University of Zurich, 8057 Zurich, Switzerland\\
$^2$Institute for Computational Science, University of Zurich, 8057 Zurich, Switzerland\\
$^3$Institute for Astronomy, Department of Physics, ETH Zurich, Wolfgang-Pauli-Strasse 27, 8093, Zurich, Switzerland\\
$^4$Kapteyn Astronomical Institute, University of Groningen, Landleven 12, Groningen NL-9747AD, The Netherlands\\
$^5$S3IT, University of Zurich, 8057 Zurich, Switzerland\\
}

\date{Submitted to MNRAS}

\pubyear{2017}

\begin{document}
\label{firstpage}
\pagerange{\pageref{firstpage}--\pageref{lastpage}}
\maketitle

\begin{abstract}

The dearth of dwarf galaxies in the local universe is hard to reconcile with the large number of low mass haloes expected within the concordance $\Lambda$CDM paradigm. Although attempts have been made to attribute the discrepancy to observational systematics, this would require an extreme modification of the density profiles of haloes through baryonic processes. In this paper we perform a systematic evaluation of the uncertainties affecting the measurement of DM halo abundance using galaxy kinematics. Using a large sample of dwarf galaxies with spatially resolved kinematic data we derive a correction to obtain the observed abundance of galaxies as a function of their halo maximum circular velocity --a direct probe of mass-- from the line-of-sight velocity function in the Local Volume. This estimate provides a direct means of comparing the predictions of theoretical models and simulations (including nonstandard cosmologies and novel galaxy formation physics) to the observational constraints. The new ``galactic $\vmax$" function is steeper than the line-of-sight velocity function but still shallower than the theoretical CDM expectation, showing that some unaccounted physical process is necessary to reduce the abundance of galaxies and/or drastically modify their density profiles compared to CDM haloes. Using this new galactic $\vmax$ function, we investigate the viability of baryonic solutions such as feedback-powered outflows and photoevaporation of gas from an ionising radiation background. However, we find that the observed relation between baryonic mass and $\Vmax$ places tight constraints on the maximum suppression from reionisation. At the 3-$\sigma$ confidence level neither energetic feedback nor photoevaporation are effective enough to reconcile the disagreement. In the case of maximum baryonic effects, the theoretical estimate still deviates significantly from the observations for $\Vmax < 60\kms$. CDM predicts at least 1.8 times more galaxies with $\vmax = 50\kms$ and 2.5 times more than observed at $30\kms$. Recent hydrodynamic simulations seem to resolve the discrepancy but disagree with the properties of observed galaxies with resolved kinematics. This abundance problem might point to the need to modify cosmological predictions at small scales.

\end{abstract}


\section{Introduction}
\label{sec:intro}

Dwarf galaxies provide a wealth of information on the formation of the smallest bound structures in the universe. They are also excellent laboratories for understanding the physics that gives rise to galaxies. However, the cosmological properties of dark matter haloes and the observed properties of the galaxies they host can be challenging to disentangle. 

A direct way to test the predictions of the cold dark matter model at small scales is to measure the observed abundance of DM haloes. However, since DM is not directly observable, we are left with observable galaxies (and perhaps  galaxy voids) as the ``peaks of the icebergs" from which to infer the abundance of their host DM haloes. This leads to several complications, including the fact that physical processes such as supernova energy release and photoheating due to reionisation may have a strong effect on the fraction of haloes that host galaxies, as well as on the detectability of these objects.

In the last two decades the cold dark matter model has been confronted with several ``small scale'' problems. First, the ``missing satellites'' refers to the underabundance of Milky Way satellite galaxies when compared to the predictions of gravity-only cosmological simulations \citep{Klypin99b,Moore99}. Photoevaporation of gas from the lowest mass haloes was presumed to be responsible for the dearth of observed satellite galaxies, although the scale and impact of this process is still under debate \citep{Gnedin00,Somerville02,Hoeft06,Nagashima06,Okamoto08}. 

More recently, new measurements of the central mass distributions of these galactic satellites allowed the CDM theory to be tested in more detail. It was found that the central velocity dispersions of the MW dwarf spheroidals are too low to host the most massive dark matter subhaloes predicted by $N$-body simulations. The so-called ``too big to fail'' problem  is more serious and difficult to solve than the missing satellites because it relates the  abundance to the structure of DM haloes using densities instead of luminosities \citep{Boylan-Kolchin11}. At a given mass or circular velocity, the cosmic abundance of haloes is predicted by cosmology to exquisite precision \citep[e.g.,][]{Springel05,Diemand07b,Klypin11,Reed13,DuttonMaccio14,Hellwing16,Heitmann16}. However, the neglected role of baryons was uncertain and better cosmological tests would require baryonic physics to be included in the models and simulations. 

Many solutions to the structure and abundance problems have been proposed since, some involving larger estimates of the mass of the MW, but most invoking a modification of the mass distribution of galactic satellite galaxies through a combination of tidal, ram-pressure stripping, and stellar feedback \citep[e.g.,][]{Arraki14,Zolotov12,Brooks13,Brooks14,Sawala16}. Although baryonic effects are now considered to be important in testing predictions of small-scale cosmology, the details and importance of each effect is difficult to assess in a self-consistent manner because of the complexity of the physics involved, and the lack of convergence in the sub-grid models used in hydrodynamics simulations.

Isolated dwarf galaxies are not subject to environmental transformations due to ram pressure and tidal stripping, and hence offer a more direct and clean way of testing cosmological predictions at small scales. $\Lambda$CDM predicts the abundance of dark matter haloes to increase steeply with decreasing maximum circular velocity. This galactic velocity function is superior to other abundance probes because it requires no assumptions about the mapping between light and mass. 

Surveys have shown that dark matter haloes expected to host galaxies are several times more numerous than observed dwarfs in the local universe. This is the so-called ``CDM dwarf overabundance problem" \citep{Zwaan10, Tikhonov09, Zavala09, Trujillo-Gomez11, Papastergis11, Klypin15, Papastergis15a, Bekeraite16}. \citet{Klypin15} recently used a volume-limited sample of galaxies in the Local Volume to infer the abundance of DM haloes assuming that the observed $\HI$ velocity width of galaxies is the same as the maximum circular velocity of their host halo. Other authors have challenged this assumption and find instead that a large correction is necessary \citep{BrookShankar16,nihaoX}. 

In this paper we develop a novel method to obtain the maximum circular velocity from the observed line-of-sight velocity width using a large complementary sample of galaxies with high-quality measurements of their spatially resolved gas kinematics, including the faintest field galaxies studied to date.  Using this correction we calculate the abundance of observed galaxies as a function of $\Vmax$. To make a direct comparison with the observations we include simple models of baryonic effects in the theoretical CDM halo velocity function.

This paper is organised as follows. Section \ref{sec:Data} describes the observational data samples and the selection criteria. In Section \ref{sec:analysis} we describe the method used to correct the observed velocity function from line-widths to halo maximum circular velocities. Section \ref{sec:results} provides functional fits to the abundance of galaxies as a function of $\vmax$, and examines the effects of feedback and reionisation on the observed DM halo VF. We discuss our results and present our conclusions in Sections \ref{sec:comparison} and \ref{sec:conclusions} respectively. Throughout the paper we assume $H_0 = 70\kms\Mpc^{-1}$ and the \emph{Planck} cosmological parameters $\Omega_{\rm m} = 0.309$, $\Omega_{\rm bar} = 0.049$, and $\sigma_8 = 0.816$.  

\section{Data and Methods}
\label{sec:Data}

In this paper we use two large and complimentary data sets of gas kinematics, stellar and $\HI$ gas mass. The first one is a very deep volume-limited sample of galaxies in the Local Volume. The Local Volume catalog contains spatially unresolved $\HI$ velocity profile widths, which do not provide the information necessary to fit DM halo density profiles. For this reason, we use in addition an extension of the \citet{PapastergisShankar16} sample of galaxies with spatially resolved $\HI$ kinematics to establish a link between the $\HI$ profile widths of the Local Volume objects and the circular velocity of their host DM haloes. We define baryonic mass as the total stellar and cold gas mass $\Mbar = \Mstar + (4/3)M_{\rm HI}$, including Helium and neglecting molecular Hydrogen. 

\subsection{Local Volume sample}
\label{sec:LVdata}

The Local Volume (LV) sample is based on the \citet{Karachentsev13} catalog. The catalog includes distances, photometry, $\HI$ mass estimates, and $\HI$ velocity widths for a volume-limited sample of galaxies with distances $D<10\Mpc$ from the Sun. The sample consists of $\sim 900$ galaxies of all morphological types and is complete down to a limiting magnitude $M_{B} = -14$, 90 per cent complete down to $M_{B} =-13.5$ (or $\vlos \simeq 20\kms$), and 50 per cent complete down to $M_B = -12$ (or $\vlos \simeq 13\kms$) \citep[][hereon K15]{Klypin15}. On average, assuming $\langle B-K\rangle= 2.35$ \citep{Jarrett03} and $M_*/L_K = 0.6$, the Local Volume sample is missing 50 per cent of all galaxies below $\Mbar \simeq 2\times 10^7\Msun$, or equivalently, below a stellar mass $\Mstar \simeq 6.3\times 10^6\Msun$. 

A subset of 620 objects have both unresolved $\HI$ 50 per cent velocity profile width ($W50$) measurements, stellar and $\HI$ masses. Most of the late-type galaxies have $W50$ data. For galaxies with $\HI$, the line-of-sight rotation velocity $\vlos$ is simply $W50/2$, while for those with no detected $\HI$ we assume the relation for dispersion dominated galaxies from K15, $\Vlos = 70 \times 10^{-(21.5+M_K)/7}\kms$ if $M_K < -15.5$, and $\Vlos = 10\kms$ otherwise. Although this estimate is more uncertain than direct $W50$ velocities, our results are unchanged if we remove these objects from the analysis since they comprise less than 10 per cent of the sample. 
 
\subsection{Spatially resolved kinematics sample}

The second data set is an extensive compilation of spatially resolved $\HI$ kinematic measurements for 202 gas rich isolated galaxies as compiled by \citet{PapastergisShankar16}. All objects have outermost-point $\HI$ deprojected rotational velocities, $\Vout = V(r_{\rm out})$. However, several objects lack a measurement of their full rotation curve, especially in the case of the lowest-mass dwarfs in our sample. All the galaxies in the sample have measurements of $\HI$ mass and stellar mass, and  178 have optical half-light radii. For those without half-light radii data, we use the fit to the relation between $r_{1/2}$ and $\Mbar$ from \citet{Bradford15}. 

In addition to resolved kinematic data, all the objects also have measured 50 per cent $\HI$ profile linewidths, allowing for direct comparison with the Local Volume data. The sample galaxies span the widest range of velocities and baryonic masses available to date for star forming galaxies, extending down to $\Mbar \sim 10^6\Msun$, which is near the completeness limit of the Local Volume sample. For further details we refer the reader to \citet{PapastergisShankar16} and references therein. From here on we refer to this sample as ``P16''.

\subsubsection{Pressure support corrections}

The $\Vout$ measurements used in this work include corrections for pressure support, so as to recover the circular velocity at the outermost radius in the $\HI$ disc, $r_{\rm out}$. For a large fraction of our sample the pressure support corrections were performed in the original references, based on the measured $\HI$ velocity dispersion and $\HI$ surface brightness profile of each object. For the 12 objects lacking a published pressure support correction, we apply a simple estimate of the form
\begin{equation}
 \Vout \rightarrow \sqrt{ \Vout^2 + 2\sigma^2 },
\end{equation}
with $\sigma = 8\kms$, following \citet{PapastergisShankar16}. This form of the so-called ``asymmetric drift'' correction assumes a radially constant velocity dispersion $\sim 8\kms$ in the outer parts of an $\HI$ disc with an exponential surface density profile and an outermost rotation measurement at a radius equal to  two disc scale-lengths. High resolution 21 cm observations of dwarfs typically find small values of the mean $\HI$ gas velocity dispersion $\sim 6 - 12\kms$ \citep{Swaters09, Oh11a, Warren12, Stilp13, Lelli14, Oh15, Iorio17}, with even lower values in the outer discs (where the gas kinematics are probed in our analysis). High resolution hydrodynamical simulations of isolated dwarfs also obtain small values for the gas velocity dispersion that are consistent with observations \citep{Read16b}. 

Pressure support corrections become most important for the lowest velocity galaxies in the sample, where gas turbulent motion becomes comparable to ordered rotation. However, the corrections are only applied to the subsample with no published measurements of velocity dispersion and, when applied, are generally small\footnote{The choice of assumptions in the asymmetric drift correction could also affect our results; however, \citet{Read16b} show that the effect on the recovered $\vmax$ is very small.}. For a galaxy with a measured $\vout = 20\kms$, the asymmetric drift correction we apply is $\sim 3\kms$, while for a dwarf with $\vout = 40\kms$ the correction becomes smaller than $2 \kms$. Assuming instead an $\HI$ dispersion $\sigma = 12\kms$ would only increase the correction to $\sim 6\kms$ for $\vout = 20\kms$, and $\sim 3\kms$ for $\vout = 40\kms$. Due to the statistical nature of our method to obtain the $\vmax$ VF, the uncertainties in asymmetric drift corrections would only affect our results if they \emph{systematically} underestimate the circular velocity. 

Table \ref{tab:table1} summarizes the velocity definitions adopted throughout the paper.

\begin{table*}
	\centering
	\caption{Summary of velocity definitions used in this work.}
	\label{tab:table1}
	\begin{tabular}{ll} 
		\hline \hline
         Symbol  & Definition \\		
		\hline
		$W50$    & \HI velocity width measured at 50 per cent of the peak flux \\
		$\vlos$  & Line-of-sight velocity obtained directly from the \HI profile width, $W50/2$ (or from \citet{Klypin15} for galaxies with no \HI, see \\
		         & Section \ref{sec:LVdata})  \\
		$\vrot$  & Deprojected line-of-sight velocity, $\vlos/(\sin i)$, where $i$ is the inclination of the galaxy. Should not be confused with the actual rotation \\ 
		         & velocity of the gas \\
		$\vout$  & Rotation velocity measured at $r_{out}$, the outermost point in the \HI rotation curve \\
		$\vcirc$ & Circular velocity of the DM halo hosting the galaxy, $\sqrt{{\rm G}M(<r)/r}$ \\
		$\vmax$  & Maximum value of the circular velocity of the halo  \\
		\hline \hline
	\end{tabular}
\end{table*}

\section{Analysis}
\label{sec:analysis}

\subsection{The galactic line-of-sight velocity function}

\begin{figure}
 \includegraphics[width=0.48\textwidth]{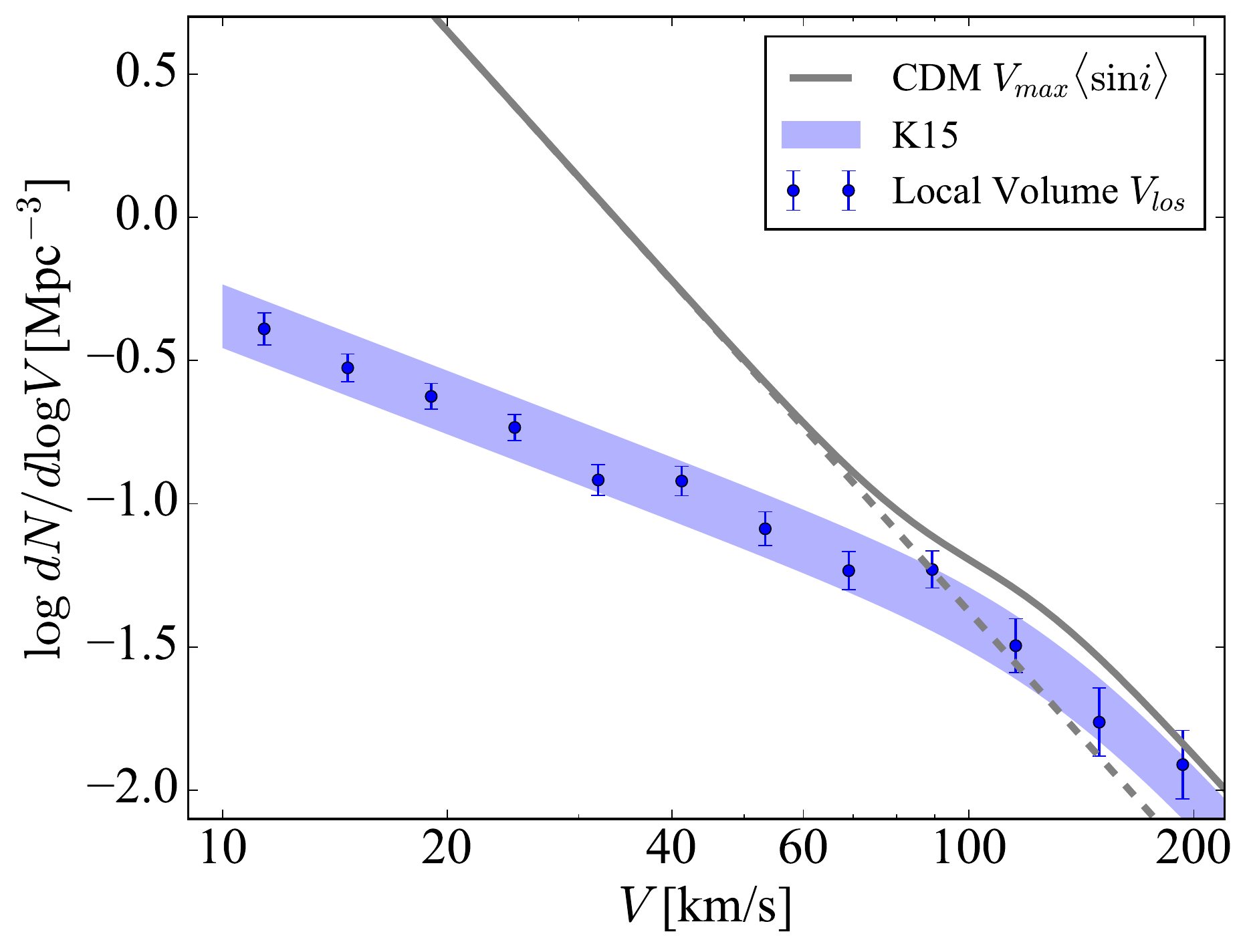} 
 \caption{Local Volume line-of-sight velocity function compared to the $\vmax$ velocity function of DM haloes predicted by CDM. The shaded area shows the fit provided by \citet{Klypin15} including statistical uncertainty. The solid curve is the fit to simulations corrected for baryon infall by \citet{Klypin15}. The dashed line is the prediction without accounting for baryon infall. The CDM $\vmax$ VF includes a $\langle \sin i \rangle$ factor to account for projection on the sky assuming random galaxy inclinations.} 
 \label{LV_VF}
\end{figure}

The starting point for our analysis is the directly measured galactic line-of-sight velocity function, which is the number density of galaxies as a function of observed $\HI$ velocity width (or line-of-sight dispersion in the case of spheroidals). Figure~\ref{LV_VF} shows the galactic line-of-sight VF in the Local Volume and the $\vmax$ VF predicted by cold dark matter structure formation. To allow a direct comparison, the halo $\Vmax$ values were projected onto the line of sight by assuming uniformly random galaxy inclinations and multiplying the analytical halo VF by the factor $\langle \sin i \rangle = \sin(60)$. The figure shows essentially the same result obtained by K15. However, K15 assumed that the correction from $\vlos$ to the halo $\vmax$ was negligible, resulting in a large discrepancy between the DM and galactic velocity functions observed in Fig.~\ref{LV_VF}. 

There are two possible ways to reconcile the observed VF with CDM. Either the rotation velocity of dwarf galaxies severely underestimates the halo maximum circular velocity, or some physical process suppresses the formation of galaxies within DM haloes below $\Vlos \sim 80\kms$. In this section we consider the first possibility. For this, we need to obtain the maximum circular velocity of each galaxy in the Local Volume by fitting density profiles to kinematic data. However, this requires resolved kinematic data from targeted observations. Most objects in the LV sample do not have resolved kinematic data so we will need to relate the LV sample line-widths with the resolved rotation velocities of the selected P16 sample. 

\begin{figure}
 \includegraphics[width=0.48\textwidth]{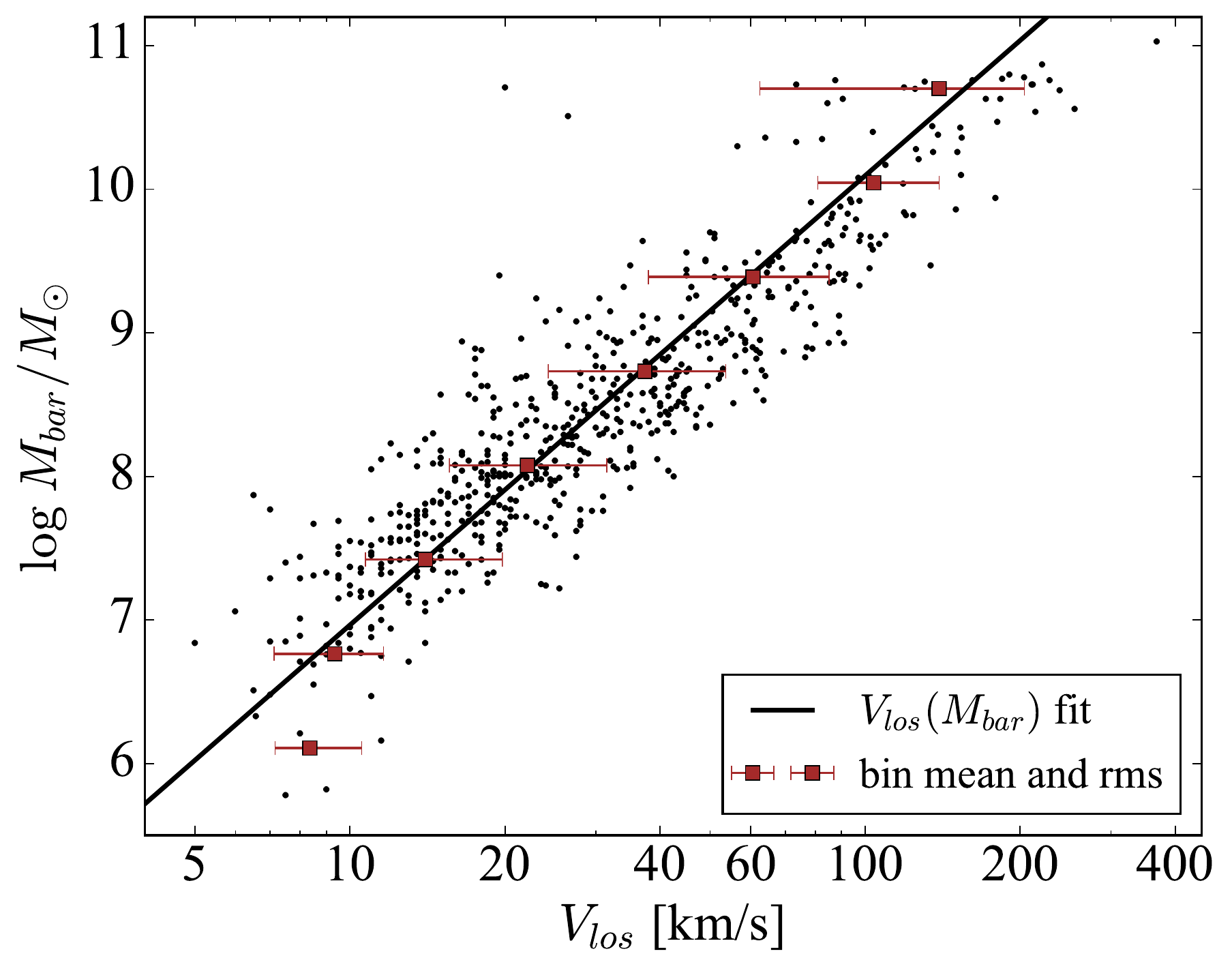} 
 \caption{Line-of-sight rotation velocity and baryonic mass for galaxies in the Local Volume sample. Squares with error bars represent the mean and standard deviation of the binned data. The line is a fit of the form $\log \Vlos(\Mbar) = \alpha \log \Mbar + \beta$, treating $\log \Mbar$ as the independent variable and neglecting individual errors. Despite the large scatter, a linear fit provides a good description of the data, except perhaps for the faintest objects where data is scarce.} 
 \label{LV_Vlos}
\end{figure}

Our objective here is to find the relation between the galaxy velocity function and the VF of the host DM haloes. Figure~\ref{LV_Vlos} shows the total baryonic mass versus $\vlos$ for the Local Volume sample. The distribution of line-of-sight velocities, correcting for incompleteness as in K15 yields the galaxy velocity function. 

To obtain an unbiased description of the relation between $\Mbar$ and $\vlos$ we calculate the distribution of the line-of-sight velocities in bins of $\log \Mbar$. Figure~\ref{LV_Vlos} shows that the resulting binned data can be appropriately described by a linear fit of the form\footnote{In this fit and throughout the paper when modelling the observed relation between $\Mbar$ and velocity we treat $\log \Mbar$ as the independent variable and velocity as the dependent variable, assuming that the velocities dominate the uncertainty. We further neglect the observational errors in the individual velocities and perform unweighted linear regressions to prevent biasing the fits towards the massive galaxies which have the smallest logarithmic velocity errors.}    
 
\begin{equation}
  \label{eq:VlosLV}
  \log \vlos^{LV} = \alpha \log \Mbar + \beta, 
\end{equation}
with $\alpha = 0.319$ and $\beta = -1.225$, and a correlation coefficient $r=0.897$. 
Having shown that the line-width baryonic Tully-Fisher relation of the LV sample is well described by a linear model, we can proceed to relate the line-of-sight $\HI$ velocity to the resolved kinematic measurements necessary for fitting dark matter halo mass profiles.

\subsection{\HI linewidths and resolved measures of galactic rotation}
\label{sec:Vlos}

Similarly to \citet{BrookShankar16}, our method relies on directly relating various galaxy kinematic measures via the observed tight coupling with total baryon mass (i.e., the Baryonic Tully-Fisher relation). The scaling between measures of gas rotation and baryonic mass has been shown to span a large range of masses from dwarfs to giant discs with very little scatter regardless of whether profile widths or spatially resolved rotation measures are used  \citep{McGaugh12, Papastergis16}. Exploiting the relatively small and well-understood systematic errors in the cold baryonic masses of isolated dwarfs (which tend to be gas dominated), we can \emph{statistically} connect the profile widths of Local Volume objects to the outermost resolved rotation measurement, $\vout$, of the P16 sample and the corresponding average $\vmax$ of their host DM haloes. 

In this section we begin with the first step in this method by quantifying the \emph{average} relation between $\HI$ linewidth-derived rotation velocity, ($\vrot$), and ($\Mbar$). This will then allow us to connect $\vrot$ to $\vmax$ using the P16 sample and profile fitting to infer the average $\vout$ - $\vmax$ relation in Section \ref{sec:profiles}. An alternative way to obtain the mean linewidth-derived rotation velocities of the LV galaxies, $\vrot$, is to assume the average deprojection for randomly sampled orientations, $\vrot = \vlos/\langle \sin i \rangle = \vlos/\sin(60)$. However, since the Local Volume catalog includes measured inclinations we opted to use their individually deprojected linewidths. 

\begin{figure}
 \includegraphics[width=0.49\textwidth]{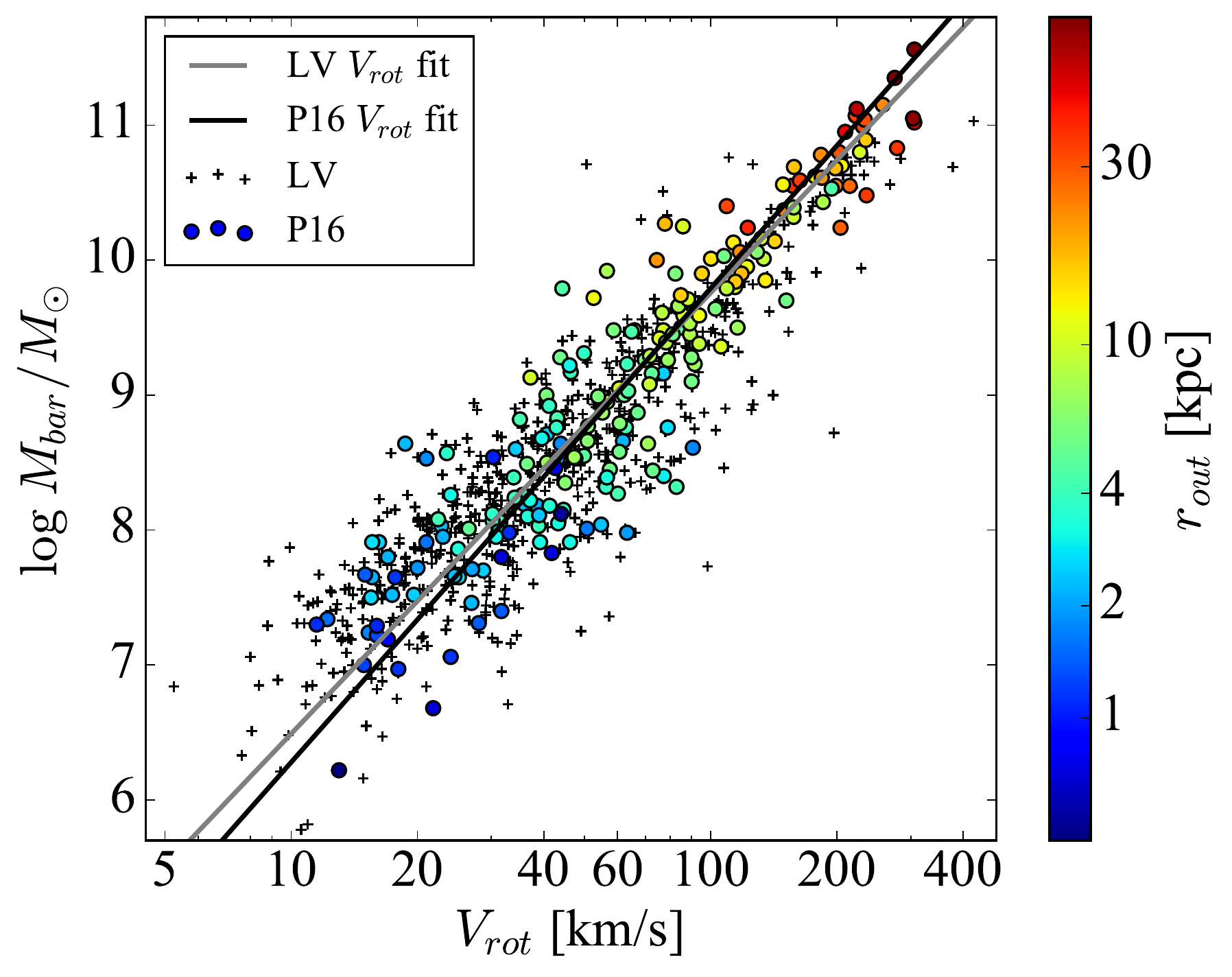} 
 \caption{Comparison of the Baryonic Tully-Fisher relation of the Local Volume and the P16 sample. For a direct comparison with the LV data we use the P16 rotation velocity obtained from the $\HI$ profile half-width, $\vrot = W50/(2\sin i)$. For P16 the colour scale shows the radius of the outermost kinematic measurement for each galaxy and the solid line shows a linear regression treating $\log \Mbar$ as the independent variable and neglecting individual errors. The plus signs represent the  LV sample and the dashed line its best linear fit. The two samples have nearly identical BTF relations while the P16 data appears to have smaller errors. This data will allow us to establish a connection between unresolved $\HI$ profile widths and spatially resolved rotation measurements in Sect. \ref{sec:profiles}. } 
 \label{LV_and_resolved}
\end{figure}

Figure~\ref{LV_and_resolved} compares the relation between baryonic mass and deprojected line-of-sight velocity, $\vrot = W50/(2\sin i)$, in the P16 and Local Volume samples. The P16 rotation velocities are well described by a linear model of the form 
\begin{equation}
\label{eq:VrotP16}
   \log \Vrot^{P16} (\Mbar) = \delta \log \Mbar + \gamma, 
\end{equation}
with $\delta = 0.289$ and $\gamma = -0.803$, and correlation coefficient $r=0.956$\footnote{For consistency, the numerical values of these parameters were obtained from a fit to the same P16 subsample used for the profile fitting in Section \ref{sec:profiles}. The two fits are almost identical within the uncertainties and this choice has a negligible effect on our results.}. We tested the linearity assumption and confirmed that power-law fits describe the data as well as non-parametric models (see Appendix \ref{sec:appendix1a}). Within the fitting uncertainties, the LV data is consistent with the same BTFR\footnote{In Appendix \ref{sec:appendix1} we show that the BTFRs of the two samples are also equivalent when inclination uncertainties are minimised by selecting only high-inclination objects.}. Here we assume that Eq. \ref{eq:VrotP16} (black line in Fig. \ref{LV_and_resolved}) represents the general galaxy population more accurately than the LV fit (grey line) due to the more precise inclination estimates obtained from the gas kinematic models used to derive rotation curves. However, this choice does not affect the results of our analysis.  

The careful reader might notice that Eq. \ref{eq:VrotP16} is different from the expected deprojection of Eq. \ref{eq:VlosLV} assuming random orientations, $\vrot = \vlos/\langle \sin i \rangle$. In general, the average $\vrot$ (at fixed $\Mbar$) follows this geometric correction for massive discs but is larger than $\vlos/\langle \sin i \rangle$ for the P16 dwarf galaxies. \citet{PapastergisShankar16} argue that this bias is due to the intrinsic thickness of the $\HI$ discs of dwarfs causing an underestimate of their true inclination in low-$i$ objects  and therefore an \emph{overestimate} of the $1/\sin i$ factor in the deprojected rotation. Only the low-$i$ dwarfs are affected by this bias, but the sample mean also shifts systematically towards higher values of $\vrot$. This explanation is confirmed by the comparison with high-inclination (and therefore more accurate) subsamples in Appendix \ref{sec:appendix1}. In our analysis we choose not to correct for this effect since it would simply shift the VF towards even lower $\vmax$ at dwarf scales, enhancing the discrepancy with CDM.  

In modelling these data we explicitly assume that for a given baryonic mass the LV and the P16 galaxies inhabit the same dark matter haloes and that systematic deviations in the observed rotation velocities are due to differences in $\HI$ content and extent that arise naturally from galaxy formation. This assumption allows us to obtain an estimate of the maximum line-width derived $\vrot$ of an LV galaxy. Using equations \ref{eq:VlosLV} and \ref{eq:VrotP16}, and setting $\langle \log \Vrot^{LV} \rangle = \langle \log \Vrot^{P16} \rangle $ as justified by Figure \ref{LV_and_resolved}, gives
\begin{equation}
    \label{eq:VrotLV}
   \langle \log \Vrot^{LV} \rangle = \langle \log \Vlos^{LV} \rangle + \log \Delta,
\end{equation}
where
\begin{equation}
    \log \Delta = (\delta - \alpha) \log \Mbar - (\gamma - \beta), 
\end{equation}
and the brackets denote population averages over narrow ranges of baryonic mass.

\subsection{Connecting $\HI$ kinematics to DM halo circular velocities}
\label{sec:profiles}

In the previous section we obtained a statistical correction to calculate the   line-width derived rotation velocity $\Vrot$ at a given line-of-sight velocity $\Vlos$ for Local Volume galaxies. The next step is to find a relation between the $\Vrot$ of the P16 galaxies and the maximum circular velocity of their host DM haloes. 

DM-only $N$-body simulations in the standard $\Lambda$CDM cosmology show that haloes have a density profile well described by the NFW parameterisation \citep{nfw97}:
\begin{equation}
  \label{NFW}
  \rho(r) = \frac{\rho_0}{\frac{r}{r_s}\left(1 + \frac{r}{r_s}\right)^2}, 
\end{equation}
with
\begin{equation}
   r_s = \Rvir/c, 
\end{equation}
and
\begin{equation}
   \Rvir = \left( \frac{3\Mvir}{4\pi \Delta_{\rm vir} \rho_{\rm m}} \right)^{1/3},
\end{equation}
where $x = r/\Rvir$, and $\Delta_{\rm vir} \approx 335$, and $\rho_{\rm m} = 1.36\times 10^2 \Omega_{\rm m} \Msun/\kpc^3$ are the virial overdensity and the matter density at present, respectively. Once the concentration is specified, the NFW profile is uniquely defined for a given $\Mvir$. To obtain the virial mass we solve the equation for the circular velocity
\begin{equation}
  \label{VNFW}
  \Vcirc(r) = V_{\rm vir}\left[ x\frac{\ln(1+cx) - \frac{cx}{1+cx}}{\ln(1+c) - \frac{c}{1+c}}\right]^{1/2},
\end{equation}
and
\begin{equation}
  V_{\rm vir} = \left(\frac{G \Mvir}{\Rvir}\right)^{1/2},
\end{equation}
where $x = r/r_s$. We solve numerically the equation\footnote{Since massive galaxies are known to have a nonnegligible baryonic contribution to their rotation curve, we subtract the enclosed baryonic mass $\Vout \rightarrow \sqrt{\Vout^2 - G\Mbar/r_{\rm out}}$ when $\Vout > 100\kms$. This assumes that the entire baryonic mass of the galaxy is contained within $r_{\rm out}$. The approximation is valid since by definition $r_{\rm out}$ occurs near the edge of the $\HI$ disc and $\HI$ is typically more extended than the stars.} 
\begin{equation}
 \label{eq:solve_vcirc}
  \Vcirc(r_{\rm out},\Mvir,c(\Mvir)) = \Vout, 
\end{equation}
for $\Mvir$, where $c(\Mvir)$ is obtained from the average concentration-mass relation \citep{DuttonMaccio14}
\begin{equation}
  \label{C-M}
  \log c(\Mvir) = 1.025\left(\frac{\Mvir}{10^{12}\Msunh}\right)^{-0.097},
\end{equation}
We then obtain the maximum circular velocity\footnote{For massive galaxies with $\Vout > 100\kms$ we add the baryon mass to the DM fit, $\Vmax = \sqrt{\Vcirc^2(r_{\rm max}) + G\Mbar/r_{\rm max}}$}
 \begin{equation}
 \label{eq:totalvcirc}
 \Vmax = \Vcirc(r_{\rm max}),
 \end{equation}
where $r_{\rm max} \approx 2.16r_s$ for the NFW profile.  For an estimate of the uncertainty in the $\Vmax$ we repeat the same calculation for the extreme values of $\Vout$ and concentration: 
\begin{equation}
  \label{errors1}
  \Vout^\pm = \Vout \pm \sigma_{\Vout}
\end{equation} 
and 
\begin{equation}
 \label{errors2}
 \log c^{\pm} = \log c(\Mvir) \mp \sigma_{\log c},
\end{equation}
where $\sigma_{\log c} = 0.11$ is the standard deviation in the concentration-mass relation \citep{DuttonMaccio14}, and $\sigma_{\Vout}$ is the reported measurement error in $\Vout$. This gives a 1-$\sigma$ upper limit to the maximum and minimum circular velocity of a halo that could host a galaxy with a given measured $\Vout$. 

In Appendix \ref{sec:appendix2} we show that the results presented here do not depend on the relative baryonic contribution to the rotation velocity, $(G\Mbar/r_{\rm out})/\Vout^2$. 
Additionally, to guarantee that the DM profiles have not been modified by core formation due to stellar feedback \citep[e.g.,][]{Mashchenko08,Pontzen12a,Governato12,nihaoIV,DiCintio14,Read16}, we further limit the sample to galaxies with $r_{\rm out} > 3 r_{1/2}$. According to \citet{Read16}, the dark matter distribution should remain unaffected by feedback at $r \gtrsim 2r_{1/2}$. About one hundred galaxies remain in the sample after this selection criteria are applied. The cuts do not significantly affect the main results of the paper (see Appendix~\ref{sec:appendix1}). 

Figure~\ref{Vout_Mbar} shows the Baryonic Tully-Fisher relation of the P16 sample using $\Vout$ as a probe of rotation velocity. In this case the correlation is also tight, with a scatter $\sigma_{\log \Vout} = 0.13$ around a linear fit. As expected, this is larger than the scatter found in other BTF samples which make stringent cuts based on the shape and extent of the rotation curves \citep[e.g.,][]{McGaugh12,Lelli16}. A fit using only galaxies in the subsample with $r_{\rm out} > 3r_{1/2}$ is essentially identical, within the uncertainties, to the full sample fit. This demonstrates that galaxies with gas discs of relatively different extent all follow the same relation between baryonic mass and rotation velocity. In other words, there is no systematic bias in the selection of the subsample used for density profile fitting. 

\begin{figure}
 \includegraphics[width=0.48\textwidth]{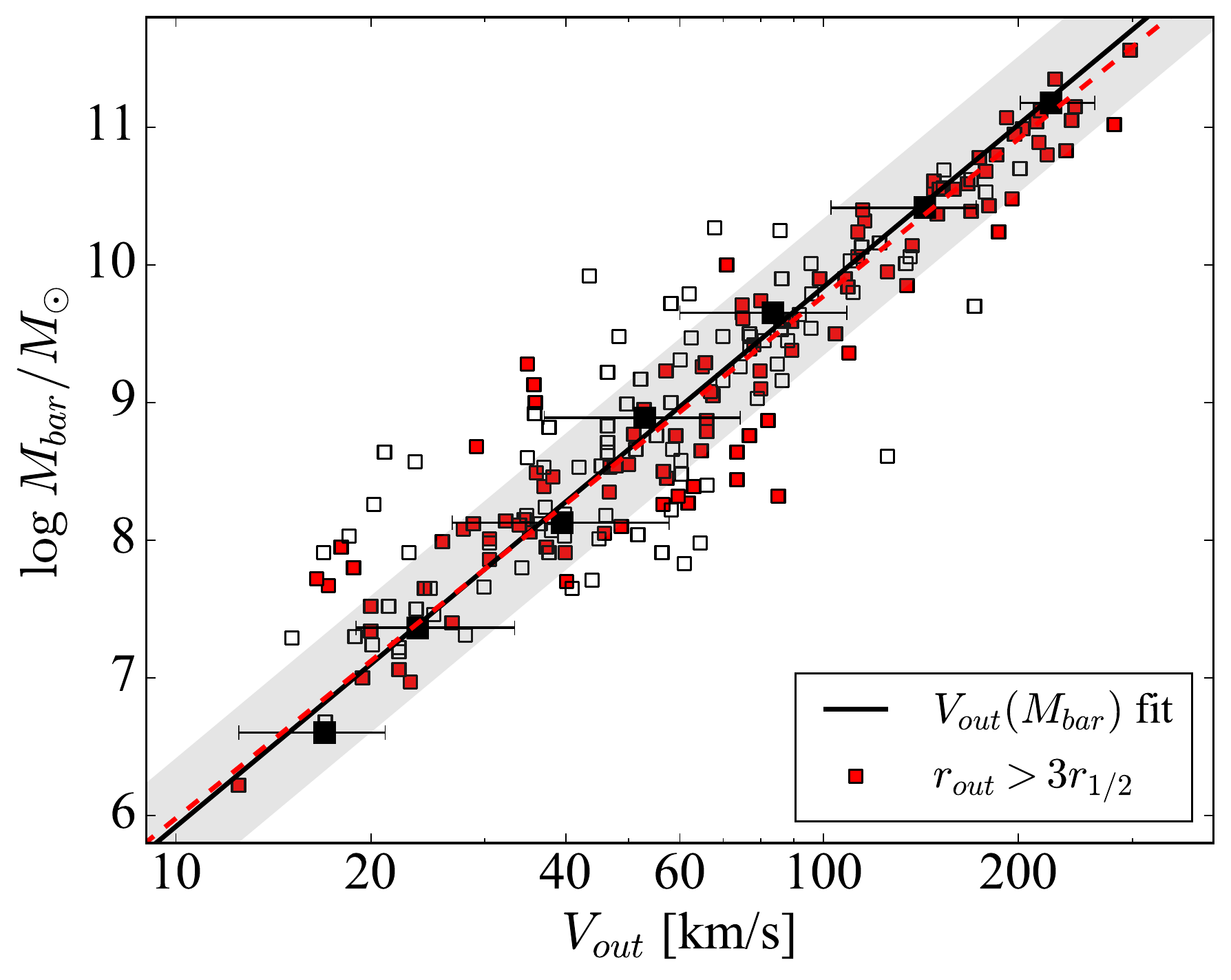} 
 \caption{Dependence of the \emph{resolved} P16 BTF relation on the relative extent of the resolved kinematic measurements. Here the rotation velocity $\vout$ is measured at the outermost kinematic radius, $r_{\rm out}$. The blue points show the selected subsample with the most extended kinematic data, $r_{\rm out} > 3r_{1/2}$. The dashed line shows a linear regression including only the selected P16 data, while the solid line and the shading show the fit to the full sample and the scatter. The points with error bars denote the binned means and scatter of the full sample in uniform logarithmic mass bins. The regressions treat $\log \Mbar$ as the independent variable and neglect individual errors. Both samples are well described by power laws. The selected sample of galaxies with the most extended kinematic data follows the same BTF relation as the complete P16 sample.} 
 \label{Vout_Mbar}
\end{figure}

\begin{figure}
 \includegraphics[width=0.49\textwidth]{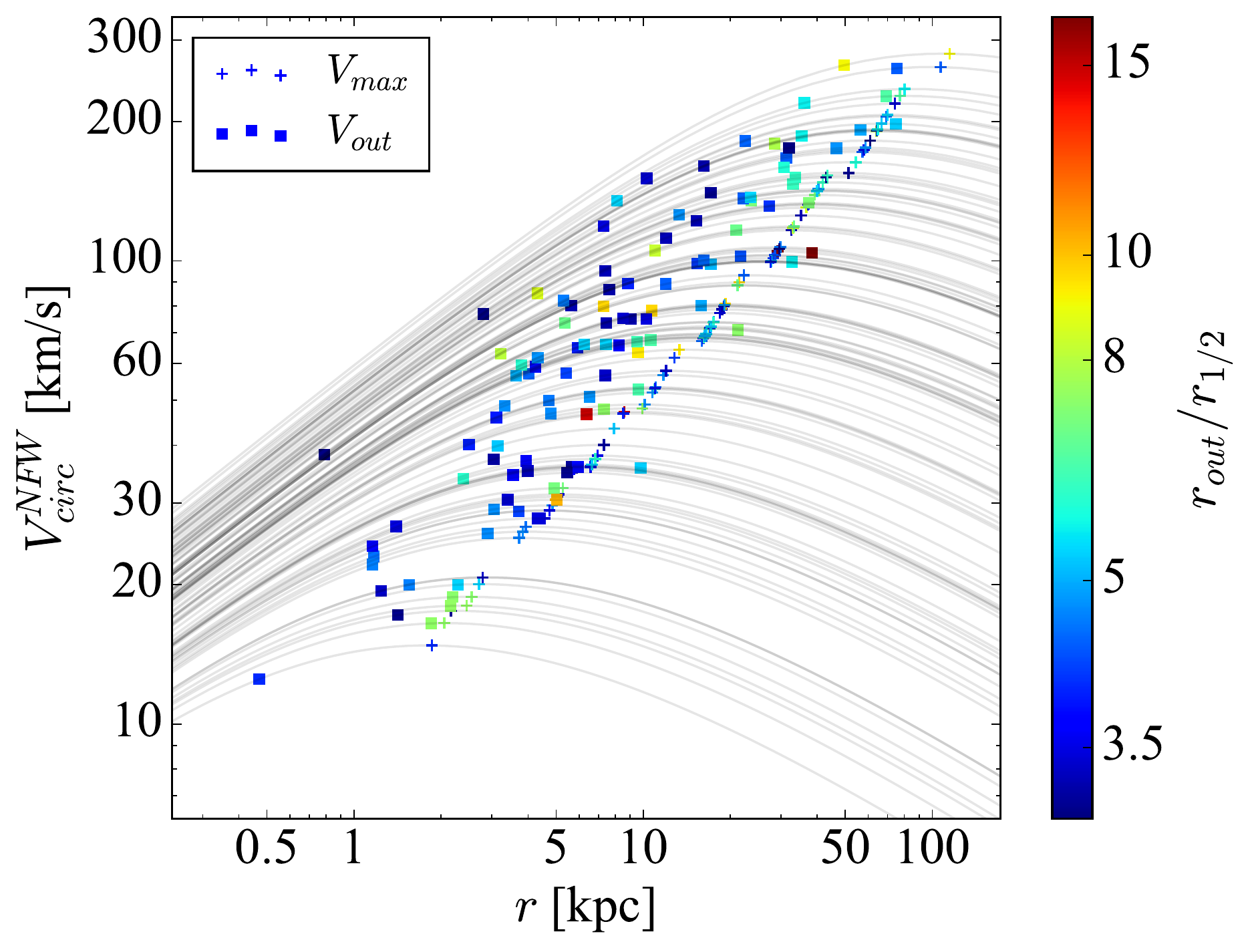} 
 \caption{NFW dark matter circular velocity profile fits to the selected P16 subsample. Solid curves show the individual profiles calculated using Eq.~\ref{VNFW}. The maximum circular velocity of the DM halo is indicated by a plus sign while the outermost resolved kinematic point is represented by a square. Error bars are omitted for clarity.} 
 \label{profiles_NFW}
\end{figure}

\begin{figure}
 \includegraphics[width=0.49\textwidth]{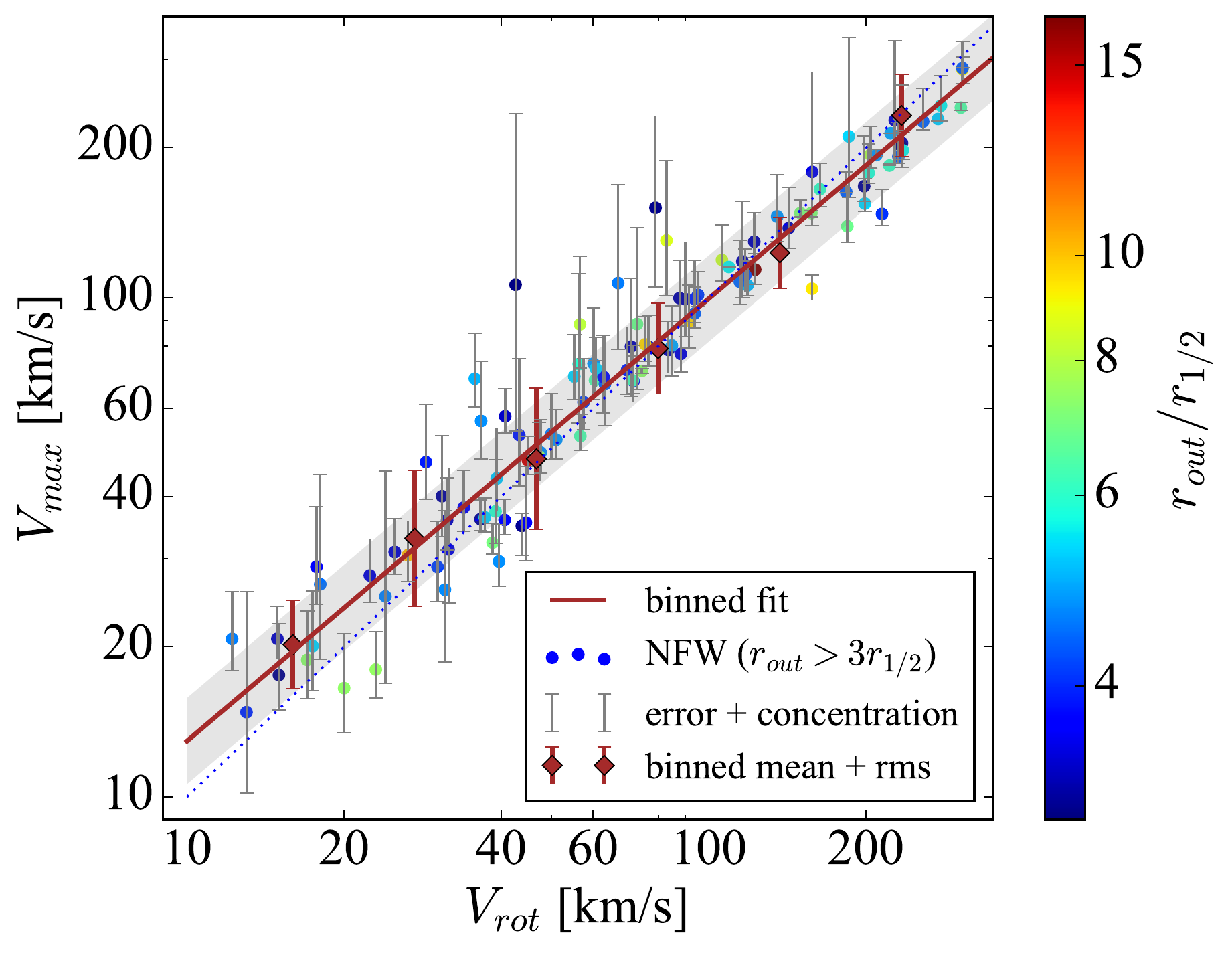} 
 \caption{Galaxy maximum circular velocity $\Vmax$ as a function of observed inclination-corrected line-of-sight velocity in the P16 selected subsample. Circles with thin error bars represent individual galaxies and their total uncertainties (due to measurement and concentration, Eqs.~\ref{errors1} and \ref{errors2}). The colour scale shows the resolved kinematic extent relative to the half-light radius. Diamonds with thick error bars denote the mean and standard deviation of the error-weighted $\Vmax$ values in uniform $\log \Vrot$ bins. The solid line is a linear fit to the binned data treating the binned error-weighted mean $\log \vmax$ values as the dependent variable and neglecting the scatter in individual bins. The shaded area shows the scatter in $\Vmax$. The dotted line indicates the relation $\Vmax = \Vrot$.} 
 \label{vrot_vnfw}
\end{figure}

The DM halo profiles obtained for the P16 subsample are shown in Figure~\ref{profiles_NFW}, and the relation between the inclination-corrected $\HI$ 50 percent half-width and $\Vmax$ is shown in Figure \ref{vrot_vnfw}. In several galaxies, the fitted circular velocity profile reaches $\Vmax$ at or near the outermost kinematic radius. However, for many, the profile keeps rising and the maximum can be at several times $r_{out}$. Figure ~\ref{profiles_NFW} shows that the difference between $\Vout$ and $\Vmax$ is in general small but not negligible. Thus, the assumption by K15 that the difference between $\Vrot$ and $\Vmax$ is less than 30 percent applies generally and agrees with our conclusions. 

In Figure \ref{vrot_vnfw} we also indicate the total uncertainties from measurement error as well as halo concentration as error bars. To avoid biases, we calculate the statistics of the error-weighted data in uniform $\log \vrot$ bins and perform a fit to the mean $\Vmax$ of the binned data. The data is well described by a linear model of the form\footnote{Here we treat $\vmax$ as the dependent variable and $\vrot$ as the independent variable and neglect the dominant uncertainties in $\vmax$ when performing the linear regression to avoid overweighting the massive galaxies. Inverting the direction of the fit has negligible effects on the result.} 
\begin{equation}
  \label{eq:vmax_vrot}
  \langle \log \Vmax \rangle =  \zeta \langle \log \Vrot^{\rm P16} \rangle + \eta,
\end{equation}
with $\zeta = 0.887$ and $\eta = 0.225$ and rms scatter $\sigma_{\Vmax} = 0.09$. As expected, massive galaxies with $\Vrot > 80\kms$ are well fit by haloes with $\Vmax$ close to the measured inclination-corrected $\HI$ profile half-width. For lower mass galaxies, the $\HI$ gas does not seem to extend far enough to probe the maximum circular velocity of the halo, resulting in a larger correction as $\Vrot$ decreases. The mean correction is less than $5\kms$ for dwarf galaxies with $\Vrot \approx 12\kms$.

The result of Figure \ref{vrot_vnfw} can now be used to re-express the observed velocity function of the Local Volume in terms of the $\Vmax$ of the haloes hosting the LV galaxies. We refer to this distribution as the ``galactic $\Vmax$ function''. Using Eqs.~\ref{eq:vmax_vrot} and \ref{eq:VrotLV} we can derive a \emph{statistical} relation between the line-of-sight rotation velocity of a galaxy in the LV and its maximum circular velocity, 
\begin{equation}
\label{eq:VmaxLV}
 \langle \log \Vmax^{\rm LV} \rangle = \langle \log \Vlos^{\rm LV} \rangle + \Lambda \langle \log\Mbar \rangle + \Psi,
\end{equation}
where
\begin{equation}
 \Lambda = \zeta\delta - \alpha = -0.0626,
\end{equation}
and
\begin{equation}
 \Psi = \zeta\gamma + \eta - \beta = 0.738,
\end{equation}
and angled brackets denote population means over narrow logarithmic $\Mbar$ bins.

\section{Results}
\label{sec:results}

\subsection{The abundance of galaxies as a function of their host halo $\vmax$}
\label{sec:VmaxVF}

Equation \ref{eq:VmaxLV} allows us to assign a $\Vmax$ to each object in the LV based on detailed modelling of the density profiles of the P16 sample galaxies. To obtain the $\vmax$ velocity function of the Local Volume we apply the following procedure:

\begin{enumerate}[1.]

 \item Using $\vlos$ and $\Mbar$ for each LV galaxy obtain $\vrot$ using Eq. \ref{eq:VrotLV}.
 
 \item Using $\vrot$ and Eq. \ref{eq:vmax_vrot} obtain $\vmax$ for each galaxy.
 
 \item Calculate the number density of LV galaxies as a function of the $\vmax$ assigned to each object including the completeness correction from K15.

\end{enumerate}

In the procedure above, steps 1-2 are equivalent to solving Eq. \ref{eq:VmaxLV} for each galaxy.  

Although we apply this correction to each object individually, the resulting $\Vmax$ is only meaningful in a statistical interpretation when an ensemble average is calculated. A caveat of this approach is that it neglects the intrinsic scatter in the BTF relations. It is  possible to repeat our analysis by modelling the scatter analytically in Eq. \ref{eq:VmaxLV}. However, our method is simpler and does not require assumptions about the error distributions. Furthermore, adding scatter to the VF does not alter its slope as long as the scatter does not depend on velocity \citep{Papastergis11}. 

\begin{figure*}
 \includegraphics[width=0.80\textwidth]{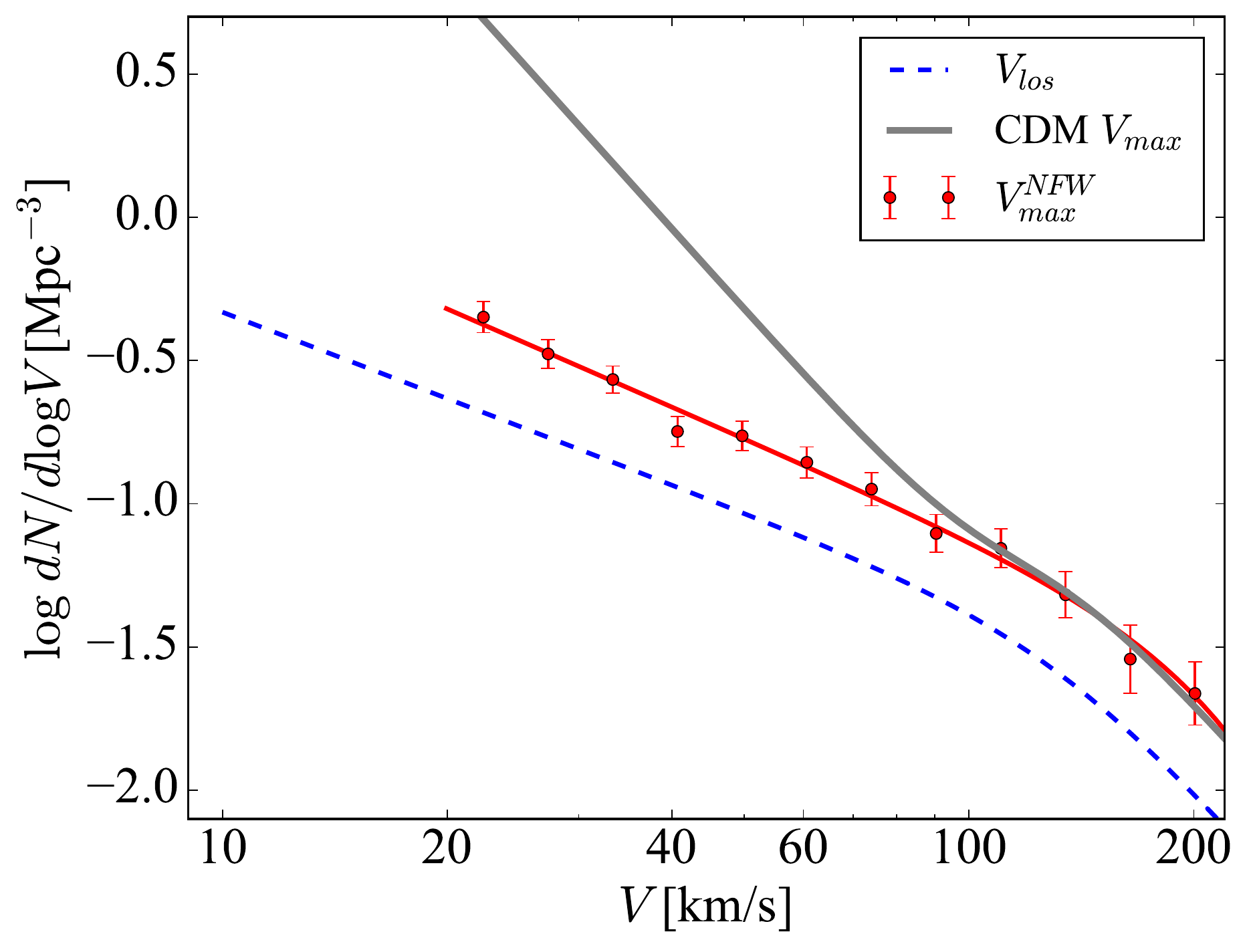} 
 \caption{Galactic $\Vmax$ function of the Local Volume. Points with error bars denote the distribution obtained by using Eq. \ref{eq:VmaxLV} to calculate the $\Vmax$ of each galaxy in the LV sample. The solid curve is a Schechter fit to the distribution (see Eq. \ref{eq:VF_fit}). The dashed curve is the observed $\Vlos$ function. The grey curve is the parametrisation of the theoretical CDM VF from K15.} 
 \label{LV_VF_obs}
\end{figure*}

Figure~\ref{LV_VF_obs} shows our main result, the $\Vmax$ velocity function in the Local Volume. The distribution is well fit by a Schechter function of the form
\begin{equation}
\label{eq:VF_fit}
 \Phi(\vmax) = \frac{dN}{d\log\Vmax} = \phi^* \left(\frac{\Vmax}{V^*}\right)^p \exp\left[-\left(\frac{\Vmax}{V^*}\right)^q \right],
\end{equation}
with $\phi^* = 2.72\times10^{-2}\Mpc^{-3}$, $V^* = 2.50\times10^{2}\kms$, $p = -1.13$, and $q = 3.14$. This fit is also shown in Figure \ref{LV_VF_obs}. The $\Vmax$ Function is slightly steeper than the observed $\Vlos$ VF but still shallower than the CDM VF because the difference between the halo $\Vmax$ and the measured rotation velocity is small and increases as galaxy mass decreases. The galactic $\Vmax$ function we obtain here should be used as a benchmark for any structure formation model to reproduce in order to be considered successful at small scales. In the next section we evaluate the effects of baryonic processes on the theoretical CDM velocity function to determine its ability to predict the abundance of small structures.

\subsubsection{The effect of feedback-induced cores on the observed $\Vmax$ function}

Galaxy formation simulations with very efficient supernova feedback implementations typically produce dwarf galaxies with DM density profiles that are shallower than NFW in the inner few kiloparsecs \citep[e.g.,][]{Mashchenko08,Governato10,Pontzen12a,Governato12,Teyssier13,DiCintio14,Trujillo-Gomez15,Onorbe15,Read16,nihaoIV}. However, the details of the transformation are still a matter of debate. For example, \citet{DiCintio14} parameterise the core-creation efficiency solely as a function of the ratio $\Mstar/\Mhalo$, while \citet{Onorbe15} and \cite{Read16} find that it also depends on the star formation history of the galaxy. The extent and slope of the DM core is also currently under debate. 

Recently, \citet{Read16} found that the effect of supernova feedback is converged once the resolution is high enough to properly capture the expansion of the blastwave. They provide a general modification of the NFW profile,
\begin{equation}
 M_{\rm coreNFW}(<r) = M_{\rm NFW}(<r) \times \left[ \tanh \left(\frac{r}{r_c}\right)\right]^n, 
\end{equation}
where
\begin{equation}
 n = \tanh \left( \kappa \frac{t_{\rm SF}}{t_{\rm dyn}} \right), 
\end{equation}
and $t_{\rm SF}$ and $t_{\rm dyn}$ are the total star formation time and the circular orbit time at the NFW scale radius respectively. Their simulations are well fit with $\kappa =0.04$. Furthermore, \citet{Read16b} show that this ``coreNFW'' profile fits ``problematic'' rotation curves within CDM. For an effectively flat core, the parameter $t_{\rm SF} >> t_{\rm dyn}$ and $n=1$. The core radius is proportional to the projected stellar half-mass radius, $r_c = 1.75 R_{1/2}$. The circular velocity of the coreNFW profile becomes
\begin{equation}
 \Vcirc^{\rm coreNFW}(r) = \Vcirc^{\rm NFW}(r) \times \left[ \tanh \left(\frac{r}{r_c}\right)\right]^{n/2}.
\end{equation}
 
We repeat the same profile-fitting procedure from Section \ref{sec:profiles} replacing Equation \ref{eq:solve_vcirc} with 
\begin{equation}
  \Vcirc^{\rm coreNFW}(r_{\rm out},\Mvir,c(\Mvir)) = \Vout, 
\end{equation}
and solving for $\Mvir$ while assuming $R_{1/2}$ is equal to the half-light radius for each galaxy. In Appendix \ref{sec:appendix3} we show that the particular choice of cored profile parameterisation has no effect on our results. 

\begin{figure}
 \includegraphics[width=0.49\textwidth]{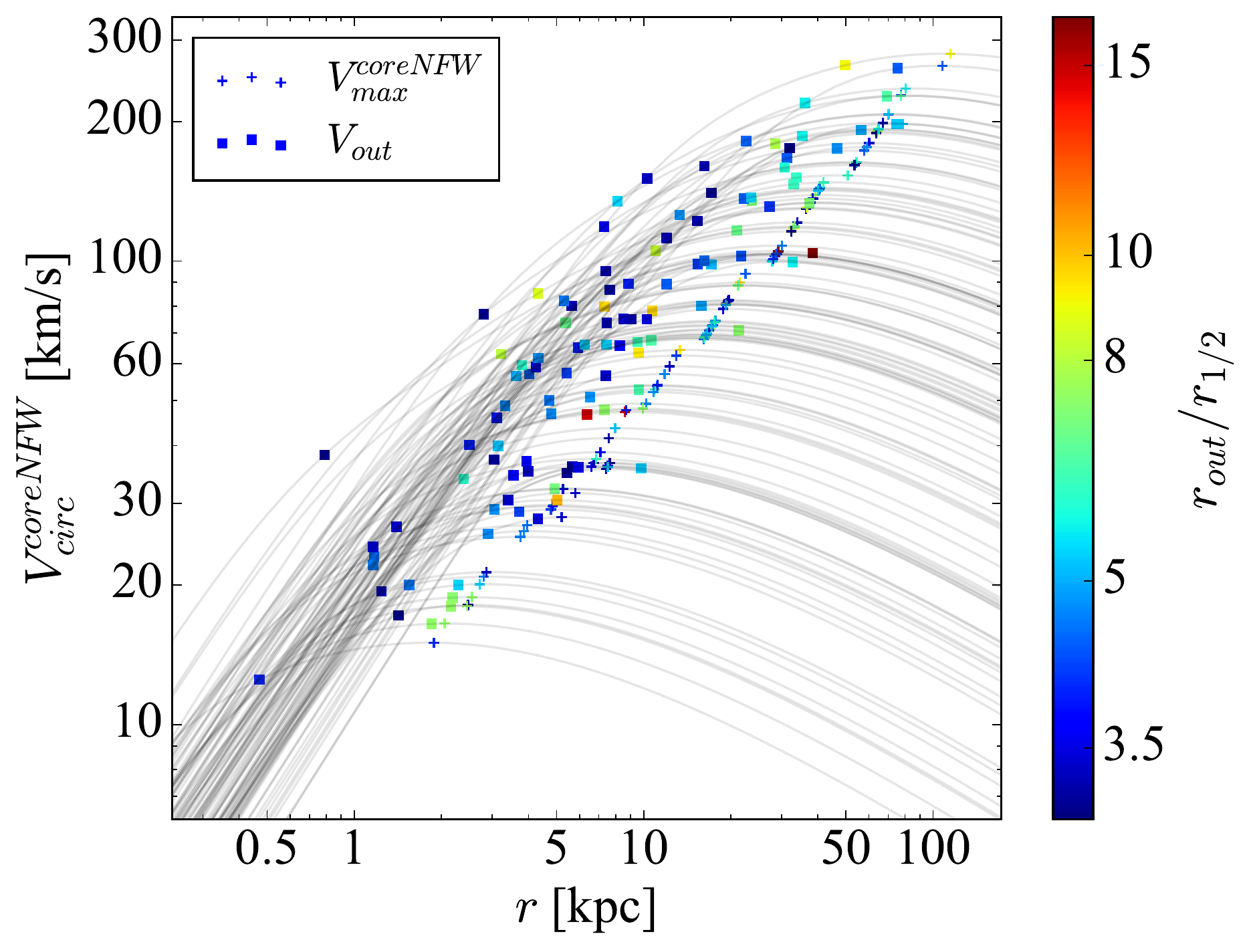} 
 \caption{Same as Fig. \ref{vrot_vnfw} but for cored NFW profiles obtained using the prescription from \citet{Read16}.} 
 \label{profiles_R16}
\end{figure}

Figure \ref{profiles_R16} shows the resulting cored DM halo fits. Although the central circular velocities are reduced with respect to Figure \ref{profiles_NFW} due to the presence of a core, the maximum circular velocity of the haloes stays relatively unchanged. This is a result of our selection of the P16 subsample with $r_{\rm out} > 3 r_{1/2}$. We emphasize here that these fits represent an extreme case where all galaxies form the most extreme shallow DM cores seen in simulations irrespective of their stellar mass. These $\Vmax$ values thus represent upper limits to the effect of feedback on the $\Vmax$ velocity function in our analysis. 

\begin{figure*}
 \includegraphics[width=0.49\textwidth]{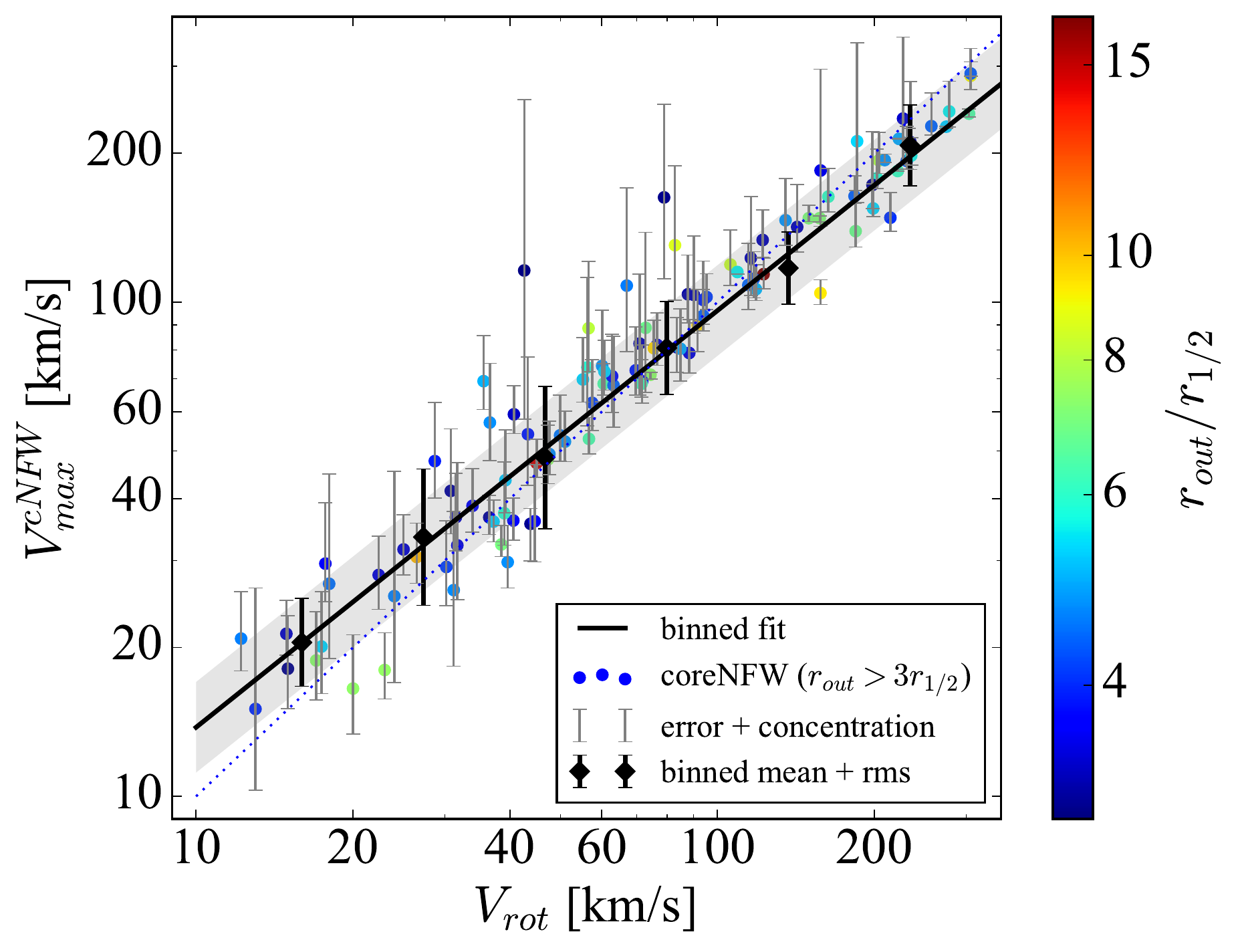} 
 \includegraphics[width=0.49\textwidth]{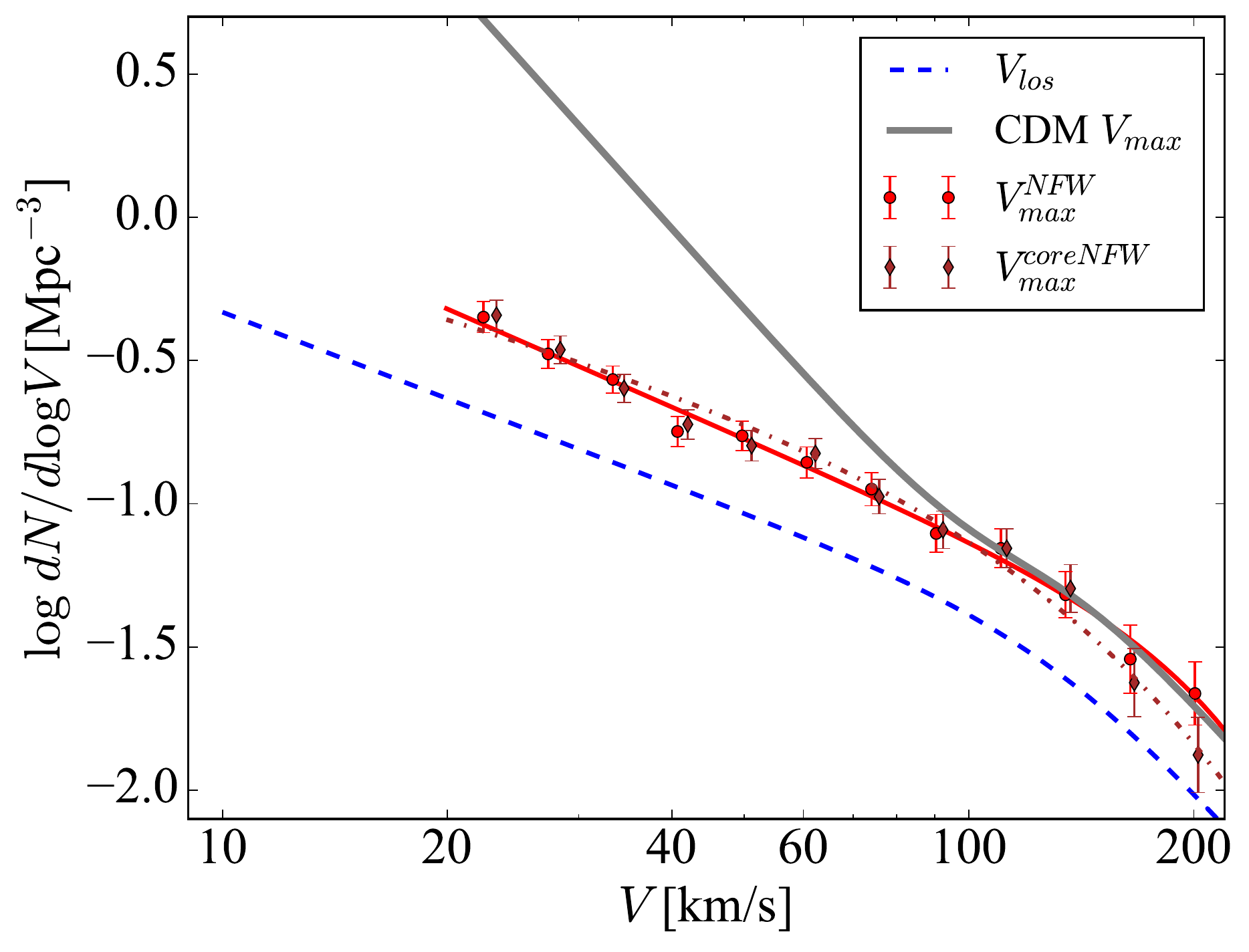} 
 \caption{Left: Same as Fig. \ref{vrot_vnfw} assuming all haloes have flat inner cores resulting from stellar feedback. The selection of galaxies with $r_{out} > 3r_{1/2}$ ensures that the core has little effect on the resulting $\Vmax$ compared to NFW profiles (see Fig. \ref{profiles_NFW}). Right: Galactic $\Vmax$ function of the Local Volume including the effects of feedback-induced cores on density profiles. Diamonds show the result of assuming maximum cores in the coreNFW parameterisation from \citet{Read16}. Squares reproduce the result in Fig. \ref{LV_VF_obs} which assumes NFW profiles. The dot-dashed and solid curves are Schechter fits to the NFW and coreNFW points respectively. The grey curve is the parameterisation of the CDM halo VF from K15.} 
 \label{vrot_vR16}
\end{figure*}

Figure \ref{vrot_vR16} shows the $\Vmax - \Vrot$ relation of the selected P16 galaxies assuming extreme feedback-induced cores. 
 
The result is essentially unchanged from Figure \ref{LV_VF_obs} because of the cut we imposed on the relative kinematic radius of the data. Ensuring that $r_{\rm out} > 3r_{1/2}$ selects galaxies for which a DM core does not modify the rotation velocity at the outermost kinematic data point. 

It should be noted that fitting cored profiles to galaxies with smaller $r_{\rm out}/r_{1/2}$ would allow some dwarfs with small kinematic radii to be placed in very massive haloes. This occurs because at a fixed circular velocity in the inner region, a cored profile will allow for a slowly rising rotation curve with a larger $\Vmax$ and $\Mvir$ \citep[see, e.g.,][]{PapastergisShankar16}. These solutions lead to extreme outliers in Fig. \ref{vrot_vR16} with large uncertainties in halo mass. We verified that most of the $\Vmax$ values of dwarfs with $r_{out} < 3r_{1/2}$ follow the relation in Fig. \ref{vrot_vR16}, while a few dwarfs get fitted to profiles with a $\Vmax$ more than twice larger than $\Vrot$. This has a negligible effect on the best fitting $\Vmax - \Vrot$ relation (Eq. \ref{eq:vmax_vrot}). Moreover, we believe the cut $r_{out} > 3r_{1/2}$ provides more reliable cosmological constraints because it probes the unmodified part of the DM halo where, in addition, baryons make a negligible mass contribution.

We have shown that observational effects and dark matter halo cores cannot account for the large discrepancy in the abundance of Local Volume galaxies compared to CDM haloes. However, theoretical work has shown that other important effects should be included when computing the theoretical abundance of galaxies hosted by DM halos. In the following section we evaluate the effects of baryon depletion due to stellar feedback and photoheating due to the reionisation of the universe.

\subsection{The impact of stellar feedback and reionisation on the observed galaxy abundance}
\label{sec:suppression}

It is expected that once the universe becomes reionised at redshift $z \sim 6$, the background ultraviolet radiation field from galaxies and quasars will have a strong effect of the formation of the faintest dwarf galaxies. In DM haloes with shallow potential wells cold neutral hydrogen will be ionised and heated. The ionised gas could then escape the halo and leave a ``dark galaxy'' behind. These dark galaxies may contain few to no stars depending on the timescales of accretion, photo-evaporation and star formation. 

Although a complete modelling of the process is extremely difficult, simplified simulations have shown that the total baryonic mass of the halo at $z=0$ is sharply suppressed for masses below a characteristic scale $\Mvir \sim 10^{9.5}\Msun$ \citep{Okamoto08}. The imprint of the transition should be detectable in galaxy samples of the smallest field dwarfs known to date. Therefore, the P16 sample is ideal to search for the signature of this process. 

\begin{figure}
 \includegraphics[width=0.48\textwidth]{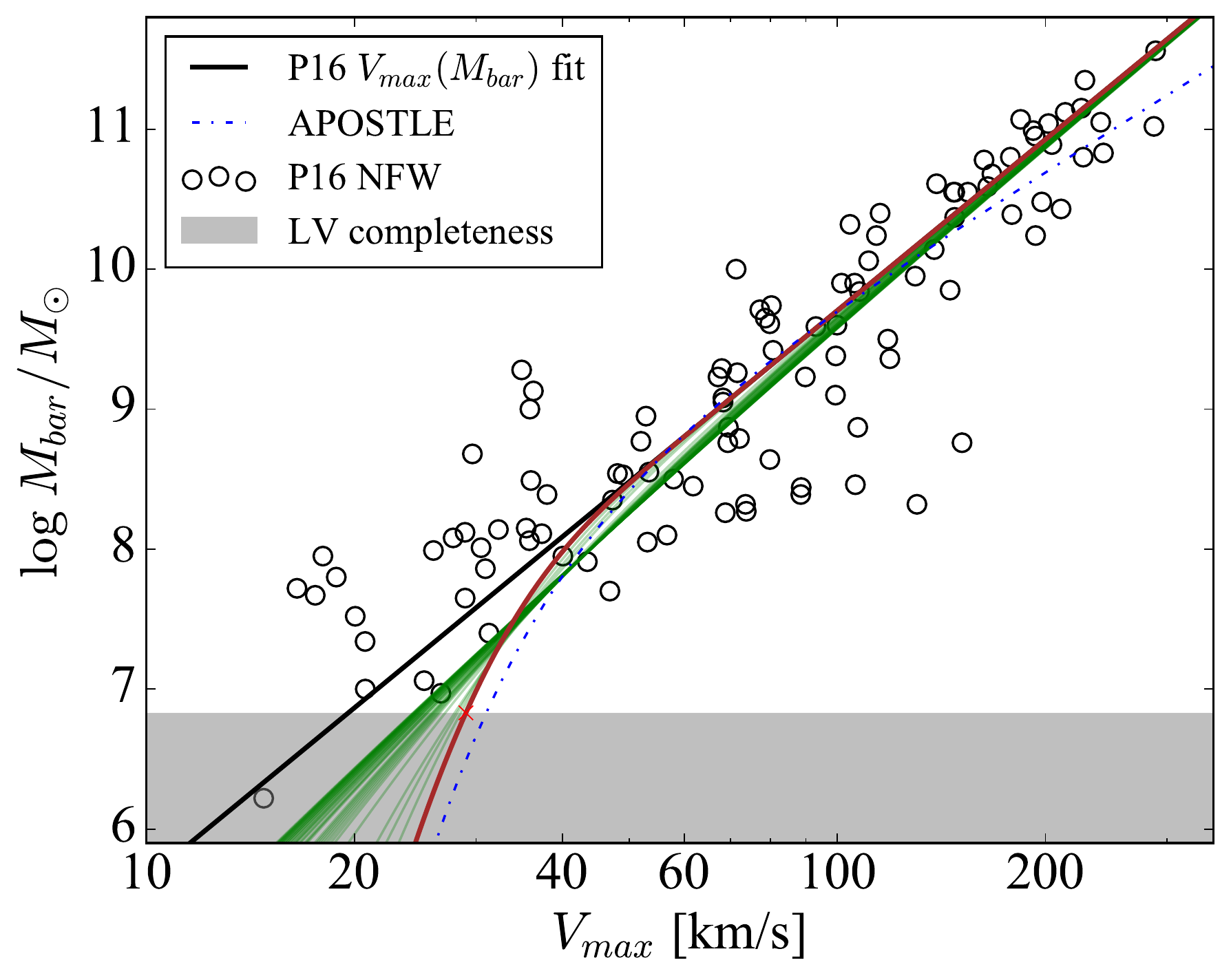} 
 \caption{Relation between cold baryonic mass and $\Vmax$ for the selected P16 galaxies. The straight line shows the no-suppression model in the form of a linear relation. The thin curves show the most extreme fits to the suppression model (Eq. \ref{eq:sup}) that are within the 3-$\sigma$ allowed region when $V_{crit}$ and $\sigma_{crit}$ are allowed to vary freely. The shaded area shows the completeness limit of the Local Volume sample. The thick curve corresponds to the suppression model with the strongest effect on the observable galaxy VF. The ``x'' marks the value $V_{crit}$ at which half of the galaxies in the LV are undetected if maximum suppression in the maximum suppression scenario. The dash-dotted line reproduces the results of the APOSTLE simulation \citep{Sales17}.} 
 \label{P16_BTF_bend}
\end{figure}

Since photo-evaporation might also affect other galaxy properties such as the extent of the $\HI$ disc, it is important to relate the baryonic mass to the depth of the potential well, using $\Vmax$. Figure \ref{P16_BTF_bend} shows $\Mbar$ versus $\Vmax$ obtained using NFW profile fitting (see Section \ref{sec:profiles}). Two processes should be dominant in setting this relation: loss of gas due to feedback-produced outflows, and photo-evaporation due to an external UV field. The physics of these processes is quite different and there is no reason to expect a simple linear scaling of the baryon mass with halo circular velocity (in logarithmic units). A second-order polynomial least squares fit of the form $\log \Vmax = a_1 \log \Mbar^2 + a_2 \log \Mbar + a_3$ yields a negligible quadratic term, indicating that the data favour a nearly perfect linear relation between maximum circular velocity and baryonic mass.

To obtain a limit on the maximum amount of baryonic loss supported by the data, we fit a no-suppression linear model of the form
\begin{equation}
 \label{eq:nosup}
 \log \vmax^{nosup} (\Mbar)= a \log \Mbar + b ,
\end{equation}
as well as model with a baryonic mass suppression term,
\begin{equation}
 \label{eq:sup}
 \Vmax^{sup}(\Mbar) = f_{sup}(\Mbar) \times \vmax^{nosup}(\Mbar),
\end{equation}
where
\begin{equation}
 f_{sup}(\vmax) =  0.5\left[1 + {\rm erf} \left( \frac{\log \vmax - \log V_{0}}{\log \sigma_{0}} \right)\right].
\end{equation}
Here, $V_{0}$ is the $\vmax$ value where the baryonic mass is reduced by 50 per cent relative to the no-suppression model, and $\sigma_{0}$ is the width of the transition. We assume that a very sharp transition is unphysical and limit the width to $\log (\sigma_0/\kms) > 1.2$, similar to the value found by \citet{Okamoto08}. Figure \ref{P16_BTF_bend} shows the results for both models. Since the model with suppression has two free parameters, we show a family of fits which are 3-$\sigma$ away from the no-suppression model using a likelihood ratio analysis. To do this, the suppression models are explored by sampling a grid of points in a 2-dimensional parameter space defined by $\sigma_0$ and $V_0$, for which the likelihood is assumed to be
\begin{equation}
    \mathcal{L} = \prod_{i=1}^{N} \frac{1}{\sqrt{2\pi\sigma^2}} \exp \left[ -\frac{ \left( V_{max,i} - \vmax^{fit}(M_{{\rm bar}, i}) \right)^2 }{ 2\sigma^2 }\right] ,
\end{equation}
where $N$ is the number of data points, $\sigma$ is the variance of the $\vmax$ data $\vmax$ with respect to the fit, and the subscript \emph{fit} refers to either the linear fit (Eq. \ref{eq:nosup}) or the fit with suppression (Eq. \ref{eq:sup}). The log-likelihood becomes
\begin{equation}
    \ln \mathcal{L} = -\frac{N}{2} \ln \sigma^2 - \frac{1}{2 \sigma^2} \sum^{N}_{i=1} \left[ V_{{\rm max},i}^2 - \vmax^{fit}(M_{{\rm bar}, i}) \right]^2 ,
\end{equation}
The logarithm of the likelihood ratio will be chi-squared distributed,
\begin{equation}
    \chi^2_{sup} = -2 \ln \frac{\mathcal{L}_{sup}}{\mathcal{L}_{nosup}} .
\end{equation}
Models with a $\chi^2_{sup}$ corresponding to a $p$-value of $0.003$ (with two degrees of freedom) are selected and shown in Figure \ref{P16_BTF_bend} as thin curves. The curve with the most extreme downward bend (thick curve) is chosen as the maximum suppression model.  

The next step to obtain the modified VF is to calculate the number of galaxies that are detected in surveys as a function of $\vmax$ and their maximum suppressed baryonic content. Assuming that the scatter in the $\Mbar - \vmax$ relation is gaussian, the detected abundance will be reduced by half at $\vmax = V_{crit}$, and the functional shape of the transition will be described by an error function of the form
\begin{equation}
\label{}
  \mathcal{F}_{ext}(\vmax) = 0.5 \left[1 + {\rm erf}\left( \frac{\log \Vmax^{sup} - \log V_{crit}}{\sqrt{2} \sigma_{\log\Vmax} }\right) \right] ,
\end{equation}
where $\Mbar^{sup}$ is given by Eq. \ref{eq:sup}, $\sigma_{\log \vmax} = 0.14$ is the logarithmic scatter in $\Vmax$ around the linear fit, and $V_{crit}$ is the value of $\Vmax$ at which 50 per cent of the galaxies would be undetectable in the Local Volume. Using the 50 per cent completeness $B$-band magnitude of the LV (see Section \ref{sec:LVdata}) gives the stellar mass completeness limit, $M_{lim} = 10^{6.8}\Msun$. To include galaxies with low gas fractions, we assume the baryonic mass completeness limit is equal to this value. Due to the steepness of the reionization downturn (see Fig. \ref{P16_BTF_bend}), the precise value of $M_{lim}$ would not significantly affect the theoretical predictions.

In addition to photo-evaporation, DM haloes hosting star formation  and energetic feedback from supernovae and stellar radiation may lose a significant fraction of their baryons through massive gas outflows \citep[e.g.,][]{Governato10,Brook11,Munshi13,Shen14,Onorbe15,Trujillo-Gomez15,nihaoI,Wheeler15}. This effect has been observed in simulations where feedback is tuned to reproduce the observed stellar mass function \citep{Sawala15}. The loss of baryons at early times reduces the accretion rate of dark matter and hence and total mass of the halo at $z=0$. This effect lowers the $\Vmax$ of all dwarf haloes and produces a net shift in the velocity function.  

The \emph{maximum} possible reduction in the mass of a halo due to internal (i.e. feedback) processes can be modelled as a reduction at high redshift in the total matter abundance by a factor equal to the baryon fraction, $\Omega_{\rm bar} \rightarrow 0$. The total matter density then becomes 
\begin{equation}
 \Omega_{\rm m}^{int} = \Omega_{\rm m} - \Omega_{\rm bar} ,
\end{equation}
where $\Omega_{\rm m}$ and $\Omega_{\rm m}$ are the \emph{Planck} total matter and baryon densities. Since the power spectrum includes a contribution from baryons, the reduction in the baryon density causes a reduction of the power given by
\begin{equation}
 P^{int}(k) = (\Omega_{\rm m}^{int}/\Omega_{\rm m})^2 P_{\rm DM}(k) = (\Omega_{\rm DM}/\Omega_{\rm m})^2 P(k).
\end{equation}
Using $\Omega_{\rm m}^{int}$ and $P^{int}(k)$ in the Extended Press-Schechter formalism \citep{Schneider13,Schneider14,Schneider15} allows us to obtain the new feedback-modified velocity function. The result can be fit with a simple reduction of the velocities,
\begin{equation}
 \Vmax^{int} = f_{int}\Vmax,
\end{equation}
with $f_{int} = 0.86$. This is equivalent to a $\sim 40$ per cent reduction in the normalisation of the VF. The total effect from external UV photo-evaporation plus internal feedback baryon depletion on the velocity function of DM haloes is then
\begin{equation}
  \Phi_{\rm gal}(\vmax) = \mathcal{F}_{ext}(\vmax^{int}/f_{int}) \times \Phi_{\rm CDM}(\vmax^{int}/f_{int}) .
\end{equation}

\begin{figure*}
 \includegraphics[width=0.80\textwidth]{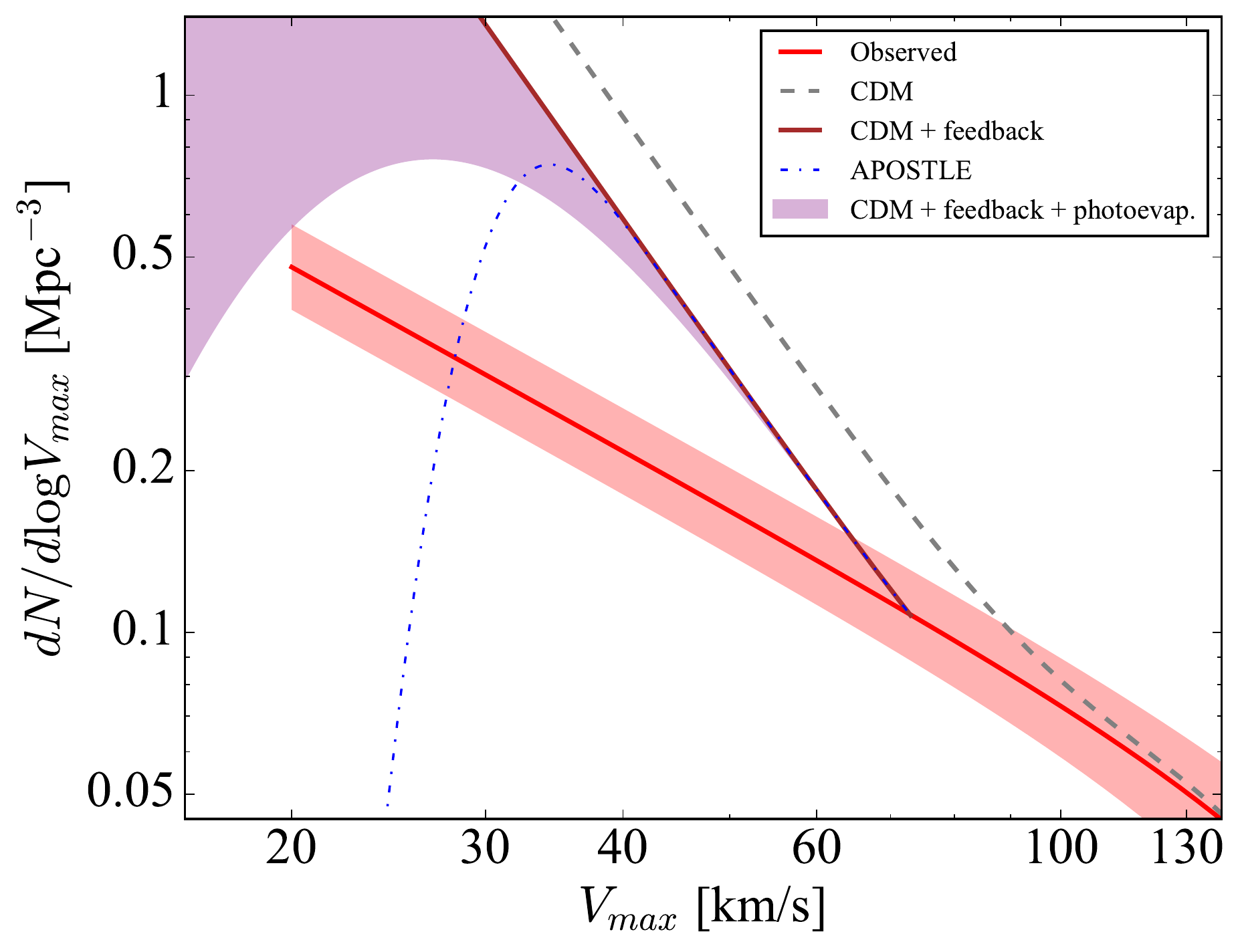} 
 \caption{Theoretical CDM velocity function including the effects of stellar feedback and photoevaporation. The dashed curve is the parameterisation of the CDM halo VF from K15. The lower solid curve includes the reduction of circular velocities of haloes due feedback-powered outflows. The shaded area includes the 3-$\sigma$ allowed region of suppression of halo detection due to photoevaporation from an ionising background (see Fig. \ref{P16_BTF_bend}). The dot-dashed curve represents the result of using the simulated APOSTLE BTF relation from Fig. \ref{P16_BTF_bend} to correct the theoretical halo velocity function for baryonic suppression effects.} 
 \label{LV_VF_obs_bar}
\end{figure*}

Figure \ref{LV_VF_obs_bar} shows the new CDM VF corrected for baryonic effects. The abundance of detected DM haloes at the lowest observed $\vmax$ is about five times lower than the original collisionless CDM estimate. This is, however, not enough to bring it into agreement with the observed galaxy $\vmax$ function in the Local Volume obtained in Section \ref{sec:VmaxVF}. Allowing for the maximum feedback and photoevaporation supression, at $\vmax = 30\kms$ the CDM galactic velocity function is still at least a factor of $\sim 2.5$ (with greater than 99.97 per cent confidence) above the observed $\vmax$ VF regardless of the assumed core/cusp nature of the density profiles. The disagreement between CDM and observations becomes significant for haloes with $\Vmax < 60\kms$, with the theory predicting $\sim 1.8$ times more galaxies than observed at $\Vmax = 50\kms$.

Figure \ref{LV_VF_obs_bar} also shows the $\vmax$ velocity function of the APOSTLE simulations derived using the $\vmax - \Mbar$ relation from \citet{Sales17} shown in Fig. \ref{P16_BTF_bend}. Although the $\vmax - \Mbar$ of APOSTLE is similar to our maximum suppression model, its smaller scatter around the relation produces a sharper cut-off in the VF (at the completeness limit of the LV).
        
\section{Comparison to other studies}
\label{sec:comparison}

\citet{BrookShankar16} performed an analysis of the rotation measurement biases in the local VF using data from the ALFALFA survey \citep{Haynes11}. They modified the theoretical CDM velocity function using abundance matching and then applied various observational BTF relations to obtain the ``observed'' ALFALFA $W50$ velocity function. They find that the definition of rotation velocity fully accounts for the large disagreement between CDM and the observed galaxy VF. 

Our conclusions are strikingly different from \citet{BrookShankar16} due to two of their key assumptions. First, the ALFALFA Baryonic Tully-Fisher relation used by \citet{BrookShankar16} is much shallower than the one we  obtained for the Local Volume (Eq. \ref{eq:VlosLV}). Second, their use of  abundance matching guarantees \emph{by construction} that any theoretical halo VF will produce the observed VF. This is simply because the mapping between halo $\vmax$ and $\Mbar$ provided by abundance matching is an ingredient of the model itself, and this ensures that the observed galaxy velocity function will always recover the \emph{input} halo VF that was assumed in deriving it (even if one assumed a non-CDM VF). Hence, an independent verification of the CDM halo velocity function would be necessary to confirm the conclusions of \citet{BrookShankar16}. We have shown in our analysis that spatially resolved dwarf kinematic data does not agree with the CDM halo VF (see Fig. \ref{vrot_vR16}).

To understand why the ALFALFA BTF relation that \citet{BrookShankar16} utilise is shallower, we repeated our analysis on the ALFALFA catalog. Using this data is more challenging because the survey is magnitude instead of volume limited.  This, combined with systematics in the faintest objects, yields a BTF relation that is poorly defined at the low-mass end. Forward fits (where $\log \Mbar$ is the dependent variable) are known to produce a bias towards shallower slopes in Tully-Fisher studies \citep[see, e.g.,][]{Bradford16} due to completeness issues at low $\Mbar$, and to the presence of outliers.

\begin{figure}
 \includegraphics[width=0.48\textwidth]{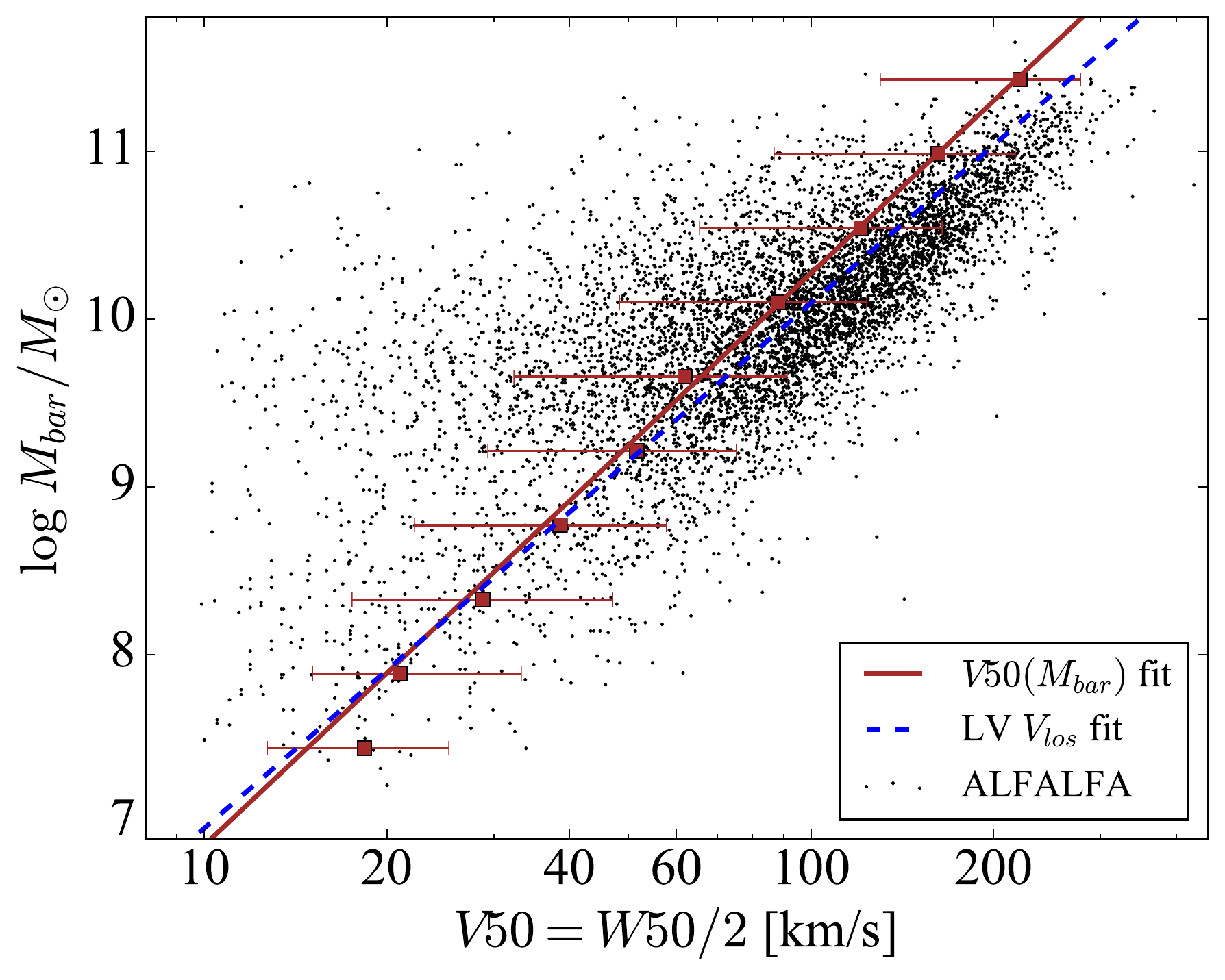} 
 \caption{ALFALFA data compared to the Local Volume sample. Points represent the ALFALFA galaxies. The solid line is a linear fit to $\log V50(\log \Mbar)$. Squares with error bars show indicate the mean and standard deviation of the data in uniform $\log\Mbar$ bins. The dashed line reproduces the fit to the LV line-of-sight velocities from Fig. \ref{LV_Vlos}. The ALFALFA fit is essentially identical to the Local Volume for $V50 < 60\kms$.} 
 \label{Mbar_V50}
\end{figure}

Figure~\ref{Mbar_V50} shows the result of performing an inverse linear fit (where $V50 = W50/2$ is the dependent variable) to the ALFALFA data. To account for deviations from a power-law, we also show the binned statistics for ten equally-spaced logarithmic bins in baryon mass. The binned fit to the ALFALFA line-of-sight velocities as a function of $\Mbar$ is comparable to the fit for the Local Volume BTF relation (Eq. \ref{eq:VlosLV}). Therefore, using the ALFALFA dataset to construct a $\vmax$ velocity function of the local universe using Eq. \ref{eq:VmaxLV} would yield the same result shown in Figure \ref{LV_VF_obs}. This indicates that the ALFALFA data is consistent with the Local Volume sample once the systematics are taken into account. 

\citet{nihaoX} used a suite of 87 galaxy formation simulations from the NIHAO project to obtain a correction from ``observed'' galaxy $W50$ to the $\Vmax$ of the host dark matter halo. These simulations explicitly include gasdynamics and metal cooling as well as standard recipes for subgrid physics such as star formation and stellar feedback. When applied to the galaxy VF, their correction is large enough to bring the observations into agreement with the CDM halo VF. To find the origin of the disagreement, we first compare the global properties of the simulations with our data and then directly compare the measurement of rotation velocity in NIHAO and the P16 data.

\begin{figure}
 \includegraphics[width=0.48\textwidth]{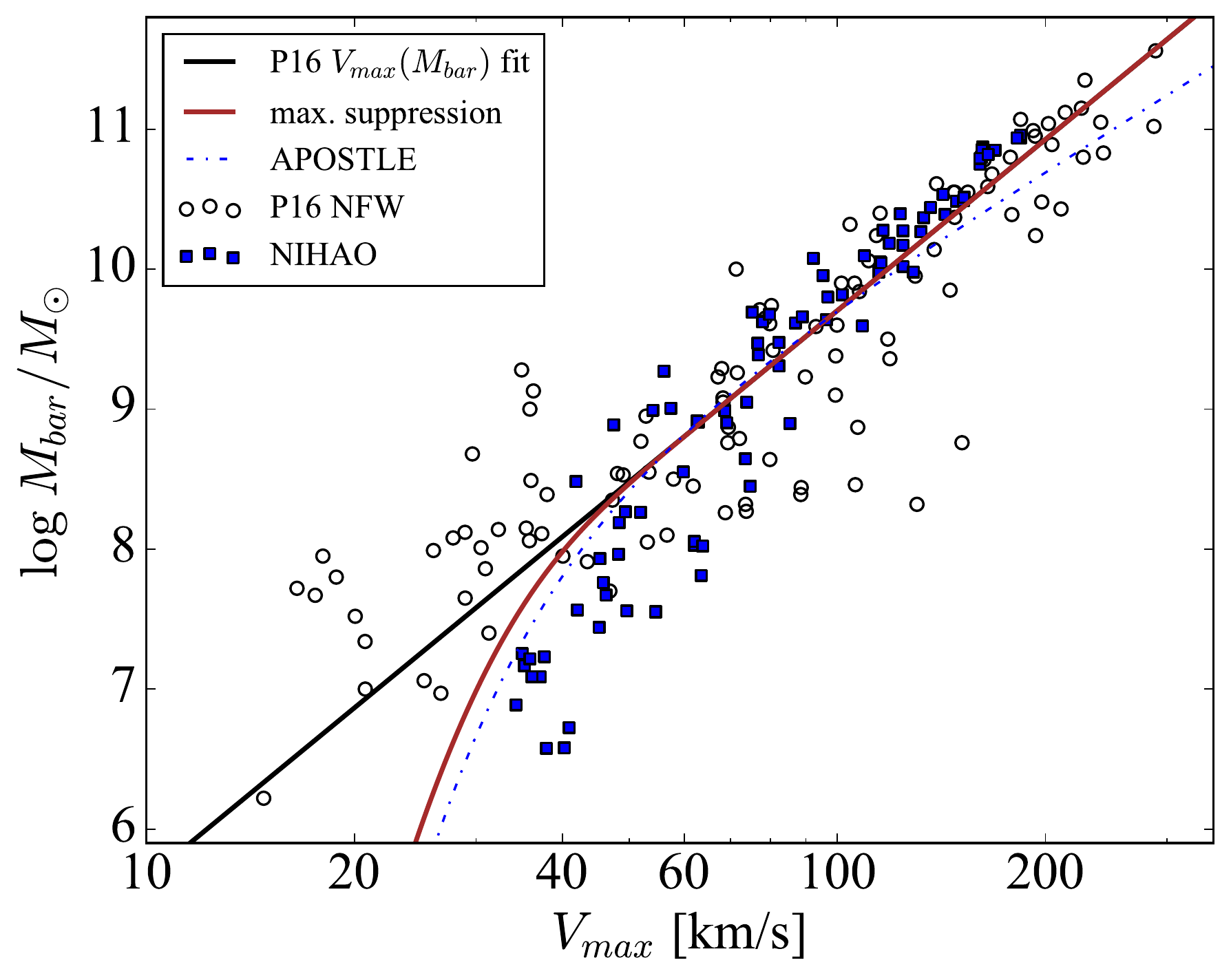} 
 \caption{Baryonic mass as a function of halo maximum circular velocity for the P16 selected sample compared to the NIHAO simulations. The APOSTLE and NIHAO simulated galaxies with $\vmax$ below $\sim 50\kms$ display a characteristic steep downturn in the relation which is not observed in the data.}
\label{nihaoBTF}
\end{figure}

Figure \ref{nihaoBTF} shows a comparison of the $\vmax$ Baryonic Tully-Fisher relation of NIHAO with the spatially resolved P16 profile fits. At a fixed baryonic mass, the NIHAO faint dwarfs inhabit haloes with larger $\vmax$ than observed galaxies. For $\Mbar < 10^8\Msun$, the discrepancy between NIHAO and  observations can be larger than a factor of $\sim 2$ in $\vmax$ or about a factor of ten in halo mass. This discrepancy is due to a sharp downturn in the baryonic mass of the NIHAO dwarfs below $\sim 50\kms$ which appears to be ruled out by the data since it lies to the right of the maximum suppression line derived from the observations in Section \ref{sec:suppression}.   

\begin{figure}
 \includegraphics[width=0.48\textwidth]{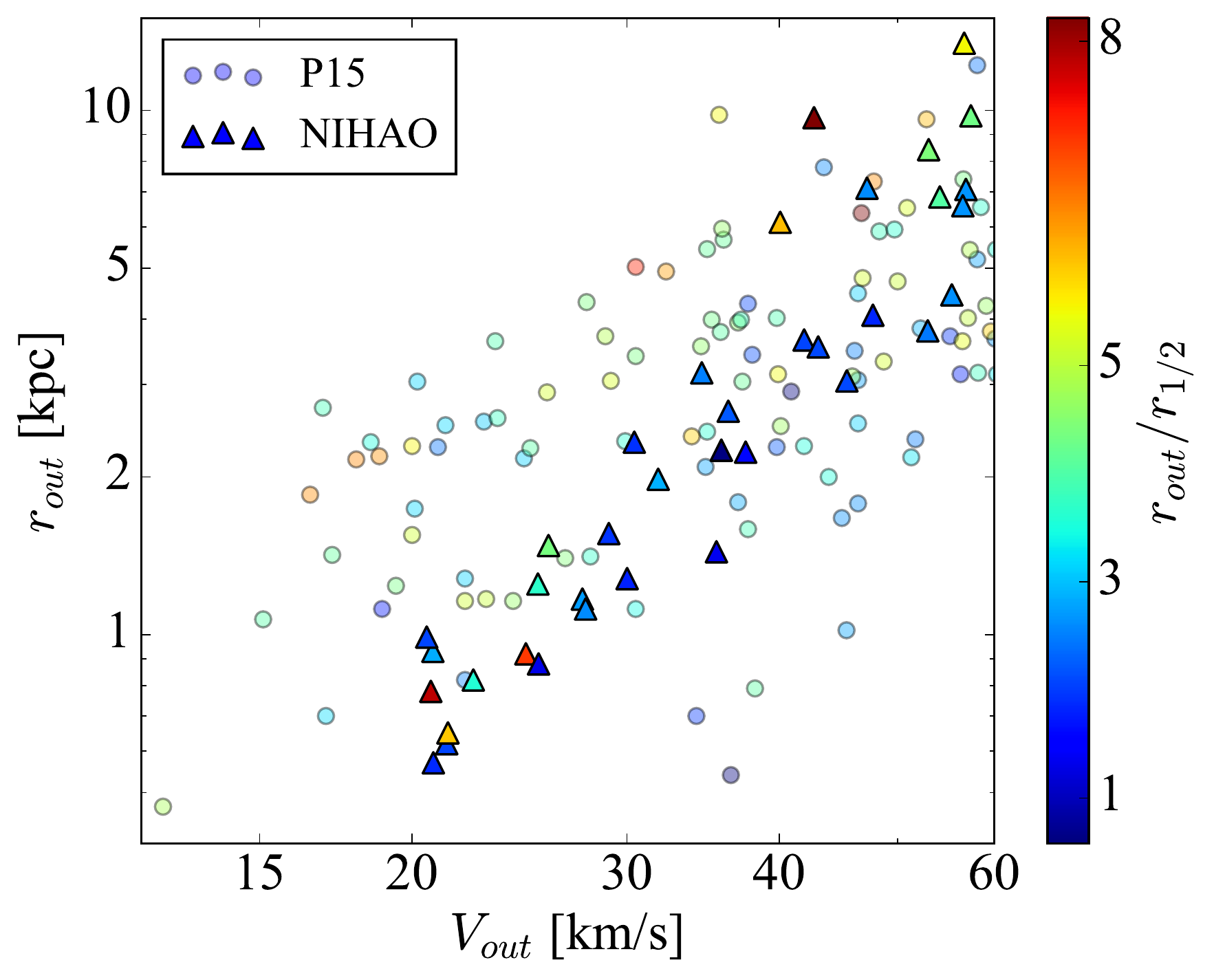} 
 \caption{Outermost kinematic radius versus $\Vout$ for the simulated NIHAO galaxies compared to the low-mass P16 observed galaxies. The colour bar shows the relative extent of the $\HI$ disc with respect to the half-light radius. For NIHAO, $\Vout$ is the circular velocity at the radius enclosing 90 per cent of the $\HI$ mass, and the projected half-light radius is measured using mock $V$-band observations. On average, observed dwarfs with $\Vrot < 40\kms$ have significantly more extended $\HI$ kinematic radii than the simulations. The NIHAO objects in this range typically have smaller discs than their observed counterparts.} 
 \label{nihao}
\end{figure}

To avoid the possibility of biases in derived quantities such as $\vmax$, we further compare the simulations with the directly observables in the spatially resolved kinematics sample. We find the following:

First, the cold gas discs of the low-mass NIHAO simulations are less extended than observed dwarfs. In Figure \ref{nihao} we show the kinematic extent of the P16 and NIHAO dwarfs as a function of their outermost $\HI$ rotation velocity. In both cases $r_{\rm out}$ is near the edge of the $\HI$ disc\footnote{In the NIHAO simulations $r_{\rm out}$ is defined as the radius enclosing 90 per cent of the $\HI$ mass.}. For objects with $\Vout \ga 40\kms$, the extent of the simulated discs seems to match the data. This is consistent with the agreement of NIHAO with the TF and BTF relations shown by \citet{nihaoXII}. However, at lower velocities there is a systematic shift in the $\HI$ kinematic extent of the simulations compared to observed galaxies. The simulations have, on average, systematically smaller cold gas discs than observed dwarfs. This may be a consequence of the strong feedback adopted in the simlations reducing the neutral gas extent at a fixed circular velocity (see Fig. \ref{nihaoBTF}).  The reduced $r_{\rm out}$ in the simulations may explain the disagreement with our VF results at $\vout \la 40\kms$. As shown in Figure \ref{vrot_vnfw}, for a galaxy with a given measured $\Vout$, the smaller $r_{\rm out}$ of NIHAO requires fitting with a more massive DM halo and a larger $\Vmax$. This, in turn, would increase the size of the correction between $\Vrot$ and $\Vmax$ shown in Figure \ref{vrot_vnfw}. Moreover, the formation of shallow dark matter cores in the NIHAO simulations \citep{nihaoIV} allows for even larger haloes to be fit when $r_{\rm out} \la r_{\rm core}$. 


Second, at a fixed $\vout$, the simulations typically show lower $\Vrot$ values than observations. In a detailed analysis, \citet{Papastergis16b} compared the full circular velocity profiles of the lowest-mass NIHAO hydrodynamic runs to the outermost rotation measurement of the P16 dwarfs with comparable $\HI$ 50 percent profile widths. They find that dwarf galaxies with large kinematic radii ($r_{\rm out} > 1.5 \kpc$) have $\Vout$ significantly below the circular velocity of the NIHAO counterparts, and conclude that these objects cannot be explained by the simulations. They also find that dwarf galaxies with large kinematic radii have, on average, larger $\Vrot$ than the NIHAO counterparts. Figure \ref{V50_Vout} indicates that the linewidth-derived $\vrot$ traces, on average, the outermost rotation velocity $\vout$ of the P16 sample to within $\sim 10$ per cent with an r.m.s. scatter of $\sim 20$ per cent.


Third and last, \citet{nihaoX} argue that compared to massive galaxies, the thicker and highly turbulent $\HI$ discs of dwarfs lead to observed $\HI$ profile widths that fail to trace the gravitational potential of the halo. Our method relies on resolved kinematic measurements which take into account turbulent support to obtain $\Vout$. This should make $\Vout$ a better tracer of the potential than projected profile widths. Nonetheless, the results of our analysis would not hold if it could be demonstrated that pressure-corrected gas rotation curves are in general a poor tracer of the mass distribution in dwarf galaxies. If this is the case in the NIHAO simulated dwarfs, it could explain the disagreement.    

Summarizing, our overall disagreement with the \citet{nihaoX} results likely originates from a combination of smaller than observed $\HI$ simulated discs in faint dwarfs, a downward bend in the $\vmax$ BTF relation which places faint galaxies into larger haloes than the data suggest, and highly turbulent simulated cold gas discs that further reduce the \HI linewidths at fixed $\vout$ in the simulations. Moreover, the most direct diagnostic, avoiding biases in derived quantities such as $\vmax$, is the relation between $W50$ and $\vout$. The $\HI$ linewidths of the NIHAO dwarf simulations are typically lower at a fixed value of $\vout$. In addition, it is also possible that the disagreement results from the failure of pressure-corrected rotation curves in accurately recovering the circular velocity (e.g., due to disequilibrium or radial motions). To examine this possibility would require mock observations of the simulated galaxies. Note, however, that \citet{Read16b} use high resolution simulations to argue against this scenario.  Another effect that might increase the discrepancy in Figure \ref{nihao} is a systematic difference in the way the extent of the $\HI$ disc is estimated. More detailed comparisons with the simulations are necessary to rule out this possibility.

\section{Conclusions}
\label{sec:conclusions}

In this paper we have performed a detailed analysis of galaxies in the Local Volume to obtain the velocity function of the dark matter haloes that host the faintest known dwarfs. We have shown that the tightly correlated Baryonic Tully-Fisher relation can be used to correct, on average, the systematic underestimation of the maximum circular velocity in kinematic data derived from spatially unresolved $\HI$ linewidths. Employing the largest available sample of dwarf galaxy spatially resolved kinematic data together with the Local Volume catalog, we obtained a \emph{statistical} relation to connect the linewidth-derived $\vrot$ to the average halo $\vmax$ as a function of $\Mbar$ by way of using parameterised mass models. This relation allowed us to derive a $\Vmax$ velocity function of galaxies for unmodified NFW as well as feedback-induced cored DM density profiles. The observed $\vmax$ VF of galaxies in the Local Volume was compared to theoretical and observational estimates of the effect of photo-evaporation and stellar feedback on the theoretical CDM halo velocity function. Our conclusions are the following:

\begin{itemize}

\item The observed $\vmax$ velocity function of galaxies is slightly steeper than the line-of-sight VF and has a higher normalisation (due mainly to inclination effects). The slightly steeper slope is a result of the $\Vmax - \Vrot$ relation (Fig. \ref{vrot_vnfw}). The new $\vmax$ VF is still well below the theoretical CDM halo VF for scales below $\Vmax \sim 80\kms$ (Fig. \ref{LV_VF_obs}).

 \item Feedback-induced dark matter cores do not significantly affect the observed $\vmax$ velocity function. This occurs because a large fraction of kinematic measurements extend well outside the region where cores are expected to form (within $\sim 2$ half-light radii), producing fits with the same halo mass (and hence $\Vmax$) as in the case with unmodified NFW profiles (Fig. \ref{vrot_vR16}). 
 
\item The maximum effect of stellar feedback on the CDM halo abundance can be estimated using a reduction of the cosmic baryon density and power spectrum amplitude at high redshift. The net effect is a reduction of the normalisation of the VF of $\sim 40$ per cent (Fig. \ref{LV_VF_obs_bar}). 

\item Photo-evaporation of gas due to reionisation can be modelled as bend in the power-law relation between baryonic mass and halo $\Vmax$ (Fig. \ref{P16_BTF_bend}). The break is not detected in current data but we can obtain an allowed (3-$\sigma$) limit on the reduction of halo detection due to reionisation suppression (Fig. \ref{P16_BTF_bend}). The data suggest that the impact of photo-evaporation from ionising radiation could become important for galaxies with $\vmax \lesssim 40\kms$. The theoretical CDM dwarf galaxy abundance is reduced by up to a factor of $\sim 2$ for halos with $\Vmax \approx 30\kms$ (Fig. \ref{LV_VF_obs_bar}). 

\item The theoretical CDM velocity function with maximum baryonic suppression is still $\sim 1.8$ times higher (at the 3-$\sigma$ level) than the observed $\Vmax$ velocity function at $\vmax = 50\kms$, and a factor of $\sim 2.5$ higher at $\vmax = 30\kms$ in the Local Volume. The discrepancy is likely to be larger if the baryonic effects are not maximal as we assumed. This discrepancy could point to the necessity of a modification of the cosmological predictions on small scales. Possible alternatives are provided by warm or self-interacting dark matter models.

\item We can compare our results to state-of-the-art galaxy formation simulations which claim to reduce the discrepancy between the observed VF and CDM. Assuming that pressure-corrected rotation curves trace, on average, the circular velocity, the disagreement results from three effects in the simulations: a steep downturn in the $\vmax$ BTF relation at $\Mbar \la 10^8\Msun$ (Fig. \ref{nihaoBTF}), a reduction in the $\HI$ extent of the lowest mass dwarfs, and an underestimation of the linewidth compared to the outermost rotation velocity in the kinematically hot simulated $\HI$ discs. The BTF downturn is disfavoured by the data (Fig. \ref{P16_BTF_bend}), while observations suggest that the mean linewidth-derived rotation velocity deviates from $\vout$ by less than $\sim 10$ per cent (Fig. \ref{V50_Vout}), and that simulated $\HI$ discs are too small (Fig. \ref{nihao}). 

\end{itemize}

Finally, we would like to note that our analysis rests on the customary assumption that extended $\HI$ rotation curve measurements (including turbulence corrections) are a good probe of the gravitational potential of the dark matter halo hosting the galaxy. If this assumption was shown to be invalid at dwarf galaxy scales, our conclusions may no longer hold. While it is extremely difficult to test the validity or accuracy of the asymmetric drift correction in dwarfs, \citet{Read16b} recently showed that in simulated galaxies with dispersions similar to observed dwarfs, the standard pressure support-corrected rotation curves yield accurate estimates of $\vcirc$. In a future paper we will further explore this issue using hydrodynamical simulations.

In the next paper in this series \citep{Schneider17} we explore the velocity functions predicted by alternative dark matter models, including baryonic effects, and compare them to the observed $\vmax$ VF obtained using the methods described here.

\section*{Acknowledgements}

We would like to thank the anonymous referee for suggesting revisions that greatly improved the quality of the paper. We also thank Anatoly Klypin for insightful discussions, and 
Andrea Macci{\`o} and Aaron Dutton for generously providing their NIHAO data. A.S. acknowledges support from the Swiss National Science Foundation (PZ00P2\_161363).



\bibliographystyle{mnras}
\bibliography{merged} 

\begin{thebibliography}{}
\makeatletter
\relax
\def\mn@urlcharsother{\let\do\@makeother \do\$\do\&\do\#\do\^\do\_\do\%\do\~}
\def\mn@doi{\begingroup\mn@urlcharsother \@ifnextchar [ {\mn@doi@}
  {\mn@doi@[]}}
\def\mn@doi@[#1]#2{\def\@tempa{#1}\ifx\@tempa\@empty \href
  {http://dx.doi.org/#2} {doi:#2}\else \href {http://dx.doi.org/#2} {#1}\fi
  \endgroup}
\def\mn@eprint#1#2{\mn@eprint@#1:#2::\@nil}
\def\mn@eprint@arXiv#1{\href {http://arxiv.org/abs/#1} {{\tt arXiv:#1}}}
\def\mn@eprint@dblp#1{\href {http://dblp.uni-trier.de/rec/bibtex/#1.xml}
  {dblp:#1}}
\def\mn@eprint@#1:#2:#3:#4\@nil{\def\@tempa {#1}\def\@tempb {#2}\def\@tempc
  {#3}\ifx \@tempc \@empty \let \@tempc \@tempb \let \@tempb \@tempa \fi \ifx
  \@tempb \@empty \def\@tempb {arXiv}\fi \@ifundefined
  {mn@eprint@\@tempb}{\@tempb:\@tempc}{\expandafter \expandafter \csname
  mn@eprint@\@tempb\endcsname \expandafter{\@tempc}}}

\bibitem[\protect\citeauthoryear{{Arraki}, {Klypin}, {More}  \&
  {Trujillo-Gomez}}{{Arraki} et~al.}{2014}]{Arraki14}
{Arraki} K.~S.,  {Klypin} A.,  {More} S.,   {Trujillo-Gomez} S.,  2014, \mn@doi
  [\mnras] {10.1093/mnras/stt2279}, \href
  {http://adsabs.harvard.edu/abs/2014MNRAS.438.1466A} {438, 1466}

\bibitem[\protect\citeauthoryear{{Bekerait{\.e}} et~al.,}{{Bekerait{\.e}}
  et~al.}{2016}]{Bekeraite16}
{Bekerait{\.e}} S.,  et~al., 2016, \mn@doi [\apjl]
  {10.3847/2041-8205/827/2/L36}, \href
  {http://adsabs.harvard.edu/abs/2016ApJ...827L..36B} {827, L36}

\bibitem[\protect\citeauthoryear{{Boylan-Kolchin}, {Bullock}  \&
  {Kaplinghat}}{{Boylan-Kolchin} et~al.}{2011}]{Boylan-Kolchin11}
{Boylan-Kolchin} M.,  {Bullock} J.~S.,   {Kaplinghat} M.,  2011, \mn@doi
  [\mnras] {10.1111/j.1745-3933.2011.01074.x}, \href
  {http://adsabs.harvard.edu/abs/2011MNRAS.415L..40B} {415, L40}

\bibitem[\protect\citeauthoryear{{Bradford}, {Geha}  \& {Blanton}}{{Bradford}
  et~al.}{2015}]{Bradford15}
{Bradford} J.~D.,  {Geha} M.~C.,   {Blanton} M.~R.,  2015, \mn@doi [\apj]
  {10.1088/0004-637X/809/2/146}, \href
  {http://adsabs.harvard.edu/abs/2015ApJ...809..146B} {809, 146}

\bibitem[\protect\citeauthoryear{{Bradford}, {Geha}  \& {van den
  Bosch}}{{Bradford} et~al.}{2016}]{Bradford16}
{Bradford} J.~D.,  {Geha} M.~C.,   {van den Bosch} F.~C.,  2016, \mn@doi [\apj]
  {10.3847/0004-637X/832/1/11}, \href
  {http://adsabs.harvard.edu/abs/2016ApJ...832...11B} {832, 11}

\bibitem[\protect\citeauthoryear{{Brook} \& {Shankar}}{{Brook} \&
  {Shankar}}{2016}]{BrookShankar16}
{Brook} C.~B.,  {Shankar} F.,  2016, \mn@doi [\mnras] {10.1093/mnras/stv2550},
  \href {http://adsabs.harvard.edu/abs/2016MNRAS.455.3841B} {455, 3841}

\bibitem[\protect\citeauthoryear{{Brook} et~al.,}{{Brook}
  et~al.}{2011}]{Brook11}
{Brook} C.~B.,  et~al., 2011, \mn@doi [\mnras]
  {10.1111/j.1365-2966.2011.18545.x}, \href
  {http://adsabs.harvard.edu/abs/2011MNRAS.415.1051B} {415, 1051}

\bibitem[\protect\citeauthoryear{{Brook}, {Stinson}, {Gibson}, {Wadsley}  \&
  {Quinn}}{{Brook} et~al.}{2012}]{Brook12}
{Brook} C.~B.,  {Stinson} G.,  {Gibson} B.~K.,  {Wadsley} J.,   {Quinn} T.,
  2012, \mn@doi [\mnras] {10.1111/j.1365-2966.2012.21306.x}, \href
  {http://adsabs.harvard.edu/abs/2012MNRAS.424.1275B} {424, 1275}

\bibitem[\protect\citeauthoryear{{Brooks} \& {Zolotov}}{{Brooks} \&
  {Zolotov}}{2014}]{Brooks14}
{Brooks} A.~M.,  {Zolotov} A.,  2014, \mn@doi [\apj]
  {10.1088/0004-637X/786/2/87}, \href
  {http://adsabs.harvard.edu/abs/2014ApJ...786...87B} {786, 87}

\bibitem[\protect\citeauthoryear{{Brooks}, {Kuhlen}, {Zolotov}  \&
  {Hooper}}{{Brooks} et~al.}{2013}]{Brooks13}
{Brooks} A.~M.,  {Kuhlen} M.,  {Zolotov} A.,   {Hooper} D.,  2013, \mn@doi
  [\apj] {10.1088/0004-637X/765/1/22}, \href
  {http://adsabs.harvard.edu/abs/2013ApJ...765...22B} {765, 22}

\bibitem[\protect\citeauthoryear{{Di Cintio}, {Brook}, {Dutton}, {Macci{\`o}},
  {Stinson}  \& {Knebe}}{{Di Cintio} et~al.}{2014}]{DiCintio14}
{Di Cintio} A.,  {Brook} C.~B.,  {Dutton} A.~A.,  {Macci{\`o}} A.~V.,
  {Stinson} G.~S.,   {Knebe} A.,  2014, \mn@doi [\mnras]
  {10.1093/mnras/stu729}, \href
  {http://adsabs.harvard.edu/abs/2014MNRAS.441.2986D} {441, 2986}

\bibitem[\protect\citeauthoryear{{Diemand}, {Kuhlen}  \& {Madau}}{{Diemand}
  et~al.}{2007}]{Diemand07b}
{Diemand} J.,  {Kuhlen} M.,   {Madau} P.,  2007, \mn@doi [\apj]
  {10.1086/520573}, \href {http://adsabs.harvard.edu/abs/2007ApJ...667..859D}
  {667, 859}

\bibitem[\protect\citeauthoryear{{Dutton} \& {Macci{\`o}}}{{Dutton} \&
  {Macci{\`o}}}{2014}]{DuttonMaccio14}
{Dutton} A.~A.,  {Macci{\`o}} A.~V.,  2014, \mn@doi [\mnras]
  {10.1093/mnras/stu742}, \href
  {http://adsabs.harvard.edu/abs/2014MNRAS.441.3359D} {441, 3359}

\bibitem[\protect\citeauthoryear{{Dutton} et~al.,}{{Dutton}
  et~al.}{2016}]{nihaoXII}
{Dutton} A.~A.,  et~al., 2016, preprint, \href
  {http://adsabs.harvard.edu/abs/2016arXiv161006375D} {} (\mn@eprint {arXiv}
  {1610.06375})

\bibitem[\protect\citeauthoryear{{Gnedin}}{{Gnedin}}{2000}]{Gnedin00}
{Gnedin} N.~Y.,  2000, \mn@doi [\apj] {10.1086/317042}, \href
  {http://adsabs.harvard.edu/abs/2000ApJ...542..535G} {542, 535}

\bibitem[\protect\citeauthoryear{{Governato} et~al.,}{{Governato}
  et~al.}{2010}]{Governato10}
{Governato} F.,  et~al., 2010, \mn@doi [\nat] {10.1038/nature08640}, \href
  {http://adsabs.harvard.edu/abs/2010Natur.463..203G} {463, 203}

\bibitem[\protect\citeauthoryear{{Governato} et~al.,}{{Governato}
  et~al.}{2012}]{Governato12}
{Governato} F.,  et~al., 2012, \mn@doi [\mnras]
  {10.1111/j.1365-2966.2012.20696.x}, \href
  {http://adsabs.harvard.edu/abs/2012MNRAS.422.1231G} {422, 1231}

\bibitem[\protect\citeauthoryear{{Haynes} et~al.,}{{Haynes}
  et~al.}{2011}]{Haynes11}
{Haynes} M.~P.,  et~al., 2011, \mn@doi [\aj] {10.1088/0004-6256/142/5/170},
  \href {http://adsabs.harvard.edu/abs/2011AJ....142..170H} {142, 170}

\bibitem[\protect\citeauthoryear{{Heitmann} et~al.,}{{Heitmann}
  et~al.}{2016}]{Heitmann16}
{Heitmann} K.,  et~al., 2016, \mn@doi [\apj] {10.3847/0004-637X/820/2/108},
  \href {http://adsabs.harvard.edu/abs/2016ApJ...820..108H} {820, 108}

\bibitem[\protect\citeauthoryear{{Hellwing}, {Frenk}, {Cautun}, {Bose},
  {Helly}, {Jenkins}, {Sawala}  \& {Cytowski}}{{Hellwing}
  et~al.}{2016}]{Hellwing16}
{Hellwing} W.~A.,  {Frenk} C.~S.,  {Cautun} M.,  {Bose} S.,  {Helly} J.,
  {Jenkins} A.,  {Sawala} T.,   {Cytowski} M.,  2016, \mn@doi [\mnras]
  {10.1093/mnras/stw214}, \href
  {http://adsabs.harvard.edu/abs/2016MNRAS.457.3492H} {457, 3492}

\bibitem[\protect\citeauthoryear{{Hoeft}, {Yepes}, {Gottl{\"o}ber}  \&
  {Springel}}{{Hoeft} et~al.}{2006}]{Hoeft06}
{Hoeft} M.,  {Yepes} G.,  {Gottl{\"o}ber} S.,   {Springel} V.,  2006, \mn@doi
  [\mnras] {10.1111/j.1365-2966.2006.10678.x}, \href
  {http://adsabs.harvard.edu/abs/2006MNRAS.371..401H} {371, 401}

\bibitem[\protect\citeauthoryear{{Iorio}, {Fraternali}, {Nipoti}, {Di Teodoro},
  {Read}  \& {Battaglia}}{{Iorio} et~al.}{2017}]{Iorio17}
{Iorio} G.,  {Fraternali} F.,  {Nipoti} C.,  {Di Teodoro} E.,  {Read} J.~I.,
  {Battaglia} G.,  2017, \mn@doi [\mnras] {10.1093/mnras/stw3285}, \href
  {http://adsabs.harvard.edu/abs/2017MNRAS.466.4159I} {466, 4159}

\bibitem[\protect\citeauthoryear{{Jarrett}, {Chester}, {Cutri}, {Schneider}  \&
  {Huchra}}{{Jarrett} et~al.}{2003}]{Jarrett03}
{Jarrett} T.~H.,  {Chester} T.,  {Cutri} R.,  {Schneider} S.~E.,   {Huchra}
  J.~P.,  2003, \mn@doi [\aj] {10.1086/345794}, \href
  {http://cdsads.u-strasbg.fr/abs/2003AJ....125..525J} {125, 525}

\bibitem[\protect\citeauthoryear{{Karachentsev}, {Makarov}  \&
  {Kaisina}}{{Karachentsev} et~al.}{2013}]{Karachentsev13}
{Karachentsev} I.~D.,  {Makarov} D.~I.,   {Kaisina} E.~I.,  2013, \mn@doi [\aj]
  {10.1088/0004-6256/145/4/101}, \href
  {http://adsabs.harvard.edu/abs/2013AJ....145..101K} {145, 101}

\bibitem[\protect\citeauthoryear{{Klypin}, {Kravtsov}, {Valenzuela}  \&
  {Prada}}{{Klypin} et~al.}{1999}]{Klypin99b}
{Klypin} A.,  {Kravtsov} A.~V.,  {Valenzuela} O.,   {Prada} F.,  1999, \mn@doi
  [\apj] {10.1086/307643}, \href
  {http://adsabs.harvard.edu/abs/1999ApJ...522...82K} {522, 82}

\bibitem[\protect\citeauthoryear{{Klypin}, {Trujillo-Gomez}  \&
  {Primack}}{{Klypin} et~al.}{2011}]{Klypin11}
{Klypin} A.~A.,  {Trujillo-Gomez} S.,   {Primack} J.,  2011, \mn@doi [\apj]
  {10.1088/0004-637X/740/2/102}, \href
  {http://adsabs.harvard.edu/abs/2011ApJ...740..102K} {740, 102}

\bibitem[\protect\citeauthoryear{{Klypin}, {Karachentsev}, {Makarov}  \&
  {Nasonova}}{{Klypin} et~al.}{2015}]{Klypin15}
{Klypin} A.,  {Karachentsev} I.,  {Makarov} D.,   {Nasonova} O.,  2015, \mn@doi
  [\mnras] {10.1093/mnras/stv2040}, \href
  {http://adsabs.harvard.edu/abs/2015MNRAS.454.1798K} {454, 1798}

\bibitem[\protect\citeauthoryear{{Lelli}, {Verheijen}  \& {Fraternali}}{{Lelli}
  et~al.}{2014}]{Lelli14}
{Lelli} F.,  {Verheijen} M.,   {Fraternali} F.,  2014, \mn@doi [\aap]
  {10.1051/0004-6361/201322657}, \href
  {http://adsabs.harvard.edu/abs/2014A%26A...566A..71L} {566, A71}

\bibitem[\protect\citeauthoryear{{Lelli}, {McGaugh}  \& {Schombert}}{{Lelli}
  et~al.}{2016}]{Lelli16}
{Lelli} F.,  {McGaugh} S.~S.,   {Schombert} J.~M.,  2016, \mn@doi [\aj]
  {10.3847/0004-6256/152/6/157}, \href
  {http://adsabs.harvard.edu/abs/2016AJ....152..157L} {152, 157}

\bibitem[\protect\citeauthoryear{{Macci{\`o}}, {Udrescu}, {Dutton}, {Obreja},
  {Wang}, {Stinson}  \& {Kang}}{{Macci{\`o}} et~al.}{2016}]{nihaoX}
{Macci{\`o}} A.~V.,  {Udrescu} S.~M.,  {Dutton} A.~A.,  {Obreja} A.,  {Wang}
  L.,  {Stinson} G.~R.,   {Kang} X.,  2016, \mn@doi [\mnras]
  {10.1093/mnrasl/slw147}, \href
  {http://adsabs.harvard.edu/abs/2016MNRAS.463L..69M} {463, L69}

\bibitem[\protect\citeauthoryear{{Mashchenko}, {Wadsley}  \&
  {Couchman}}{{Mashchenko} et~al.}{2008}]{Mashchenko08}
{Mashchenko} S.,  {Wadsley} J.,   {Couchman} H.~M.~P.,  2008, \mn@doi [Science]
  {10.1126/science.1148666}, \href
  {http://adsabs.harvard.edu/abs/2008Sci...319..174M} {319, 174}

\bibitem[\protect\citeauthoryear{{McGaugh}}{{McGaugh}}{2012}]{McGaugh12}
{McGaugh} S.~S.,  2012, \mn@doi [\aj] {10.1088/0004-6256/143/2/40}, \href
  {http://adsabs.harvard.edu/abs/2012AJ....143...40M} {143, 40}

\bibitem[\protect\citeauthoryear{{Moore}, {Ghigna}, {Governato}, {Lake},
  {Quinn}, {Stadel}  \& {Tozzi}}{{Moore} et~al.}{1999}]{Moore99}
{Moore} B.,  {Ghigna} S.,  {Governato} F.,  {Lake} G.,  {Quinn} T.,  {Stadel}
  J.,   {Tozzi} P.,  1999, \mn@doi [\apjl] {10.1086/312287}, \href
  {http://adsabs.harvard.edu/abs/1999ApJ...524L..19M} {524, L19}

\bibitem[\protect\citeauthoryear{{Munshi} et~al.,}{{Munshi}
  et~al.}{2013}]{Munshi13}
{Munshi} F.,  et~al., 2013, \mn@doi [\apj] {10.1088/0004-637X/766/1/56}, \href
  {http://adsabs.harvard.edu/abs/2013ApJ...766...56M} {766, 56}

\bibitem[\protect\citeauthoryear{{Nagashima} \& {Okamoto}}{{Nagashima} \&
  {Okamoto}}{2006}]{Nagashima06}
{Nagashima} M.,  {Okamoto} T.,  2006, \mn@doi [\apj] {10.1086/503180}, \href
  {http://adsabs.harvard.edu/abs/2006ApJ...643..863N} {643, 863}

\bibitem[\protect\citeauthoryear{{Navarro}, {Frenk}  \& {White}}{{Navarro}
  et~al.}{1997}]{nfw97}
{Navarro} J.~F.,  {Frenk} C.~S.,   {White} S.~D.~M.,  1997, \apj, \href
  {http://adsabs.harvard.edu/cgi-bin/nph-bib_query?bibcode=1997ApJ...490..493N&db_key=AST}
  {490, 493}

\bibitem[\protect\citeauthoryear{{O{\~n}orbe}, {Boylan-Kolchin}, {Bullock},
  {Hopkins}, {Kere{\v s}}, {Faucher-Gigu{\`e}re}, {Quataert}  \&
  {Murray}}{{O{\~n}orbe} et~al.}{2015}]{Onorbe15}
{O{\~n}orbe} J.,  {Boylan-Kolchin} M.,  {Bullock} J.~S.,  {Hopkins} P.~F.,
  {Kere{\v s}} D.,  {Faucher-Gigu{\`e}re} C.-A.,  {Quataert} E.,   {Murray} N.,
   2015, \mn@doi [\mnras] {10.1093/mnras/stv2072}, \href
  {http://adsabs.harvard.edu/abs/2015MNRAS.454.2092O} {454, 2092}

\bibitem[\protect\citeauthoryear{{Oh}, {de Blok}, {Brinks}, {Walter}  \&
  {Kennicutt}}{{Oh} et~al.}{2011}]{Oh11a}
{Oh} S.-H.,  {de Blok} W.~J.~G.,  {Brinks} E.,  {Walter} F.,   {Kennicutt} Jr.
  R.~C.,  2011, \mn@doi [\aj] {10.1088/0004-6256/141/6/193}, \href
  {http://adsabs.harvard.edu/abs/2011AJ....141..193O} {141, 193}

\bibitem[\protect\citeauthoryear{{Oh} et~al.,}{{Oh} et~al.}{2015}]{Oh15}
{Oh} S.-H.,  et~al., 2015, \mn@doi [\aj] {10.1088/0004-6256/149/6/180}, \href
  {http://adsabs.harvard.edu/abs/2015AJ....149..180O} {149, 180}

\bibitem[\protect\citeauthoryear{{Okamoto}, {Gao}  \& {Theuns}}{{Okamoto}
  et~al.}{2008}]{Okamoto08}
{Okamoto} T.,  {Gao} L.,   {Theuns} T.,  2008, \mn@doi [\mnras]
  {10.1111/j.1365-2966.2008.13830.x}, \href
  {http://adsabs.harvard.edu/abs/2008MNRAS.390..920O} {390, 920}

\bibitem[\protect\citeauthoryear{{Papastergis} \& {Ponomareva}}{{Papastergis}
  \& {Ponomareva}}{2016}]{Papastergis16b}
{Papastergis} E.,  {Ponomareva} A.~A.,  2016, preprint, \href
  {http://adsabs.harvard.edu/abs/2016arXiv160805214P} {} (\mn@eprint {arXiv}
  {1608.05214})

\bibitem[\protect\citeauthoryear{{Papastergis} \& {Shankar}}{{Papastergis} \&
  {Shankar}}{2016}]{PapastergisShankar16}
{Papastergis} E.,  {Shankar} F.,  2016, \mn@doi [\aap]
  {10.1051/0004-6361/201527854}, \href
  {http://adsabs.harvard.edu/abs/2016A%26A...591A..58P} {591, A58}

\bibitem[\protect\citeauthoryear{{Papastergis}, {Martin}, {Giovanelli}  \&
  {Haynes}}{{Papastergis} et~al.}{2011}]{Papastergis11}
{Papastergis} E.,  {Martin} A.~M.,  {Giovanelli} R.,   {Haynes} M.~P.,  2011,
  \mn@doi [\apj] {10.1088/0004-637X/739/1/38}, \href
  {http://adsabs.harvard.edu/abs/2011ApJ...739...38P} {739, 38}

\bibitem[\protect\citeauthoryear{{Papastergis}, {Giovanelli}, {Haynes}  \&
  {Shankar}}{{Papastergis} et~al.}{2015}]{Papastergis15a}
{Papastergis} E.,  {Giovanelli} R.,  {Haynes} M.~P.,   {Shankar} F.,  2015,
  \mn@doi [\aap] {10.1051/0004-6361/201424909}, \href
  {http://adsabs.harvard.edu/abs/2015A%26A...574A.113P} {574, A113}

\bibitem[\protect\citeauthoryear{{Papastergis}, {Adams}  \& {van der
  Hulst}}{{Papastergis} et~al.}{2016}]{Papastergis16}
{Papastergis} E.,  {Adams} E.~A.~K.,   {van der Hulst} J.~M.,  2016, preprint,
  \href {http://adsabs.harvard.edu/abs/2016arXiv160209087P} {} (\mn@eprint
  {arXiv} {1602.09087})

\bibitem[\protect\citeauthoryear{{Pontzen} \& {Governato}}{{Pontzen} \&
  {Governato}}{2012}]{Pontzen12a}
{Pontzen} A.,  {Governato} F.,  2012, \mn@doi [\mnras]
  {10.1111/j.1365-2966.2012.20571.x}, \href
  {http://adsabs.harvard.edu/abs/2012MNRAS.421.3464P} {421, 3464}

\bibitem[\protect\citeauthoryear{{Read}, {Agertz}  \& {Collins}}{{Read}
  et~al.}{2016a}]{Read16}
{Read} J.~I.,  {Agertz} O.,   {Collins} M.~L.~M.,  2016a, \mn@doi [\mnras]
  {10.1093/mnras/stw713}, \href
  {http://adsabs.harvard.edu/abs/2016MNRAS.459.2573R} {459, 2573}

\bibitem[\protect\citeauthoryear{{Read}, {Iorio}, {Agertz}  \&
  {Fraternali}}{{Read} et~al.}{2016b}]{Read16b}
{Read} J.~I.,  {Iorio} G.,  {Agertz} O.,   {Fraternali} F.,  2016b, \mn@doi
  [\mnras] {10.1093/mnras/stw1876}, \href
  {http://adsabs.harvard.edu/abs/2016MNRAS.462.3628R} {462, 3628}

\bibitem[\protect\citeauthoryear{{Reed}, {Smith}, {Potter}, {Schneider},
  {Stadel}  \& {Moore}}{{Reed} et~al.}{2013}]{Reed13}
{Reed} D.~S.,  {Smith} R.~E.,  {Potter} D.,  {Schneider} A.,  {Stadel} J.,
  {Moore} B.,  2013, \mn@doi [\mnras] {10.1093/mnras/stt301}, \href
  {http://adsabs.harvard.edu/abs/2013MNRAS.431.1866R} {431, 1866}

\bibitem[\protect\citeauthoryear{{Sales} et~al.,}{{Sales}
  et~al.}{2017}]{Sales17}
{Sales} L.~V.,  et~al., 2017, \mn@doi [\mnras] {10.1093/mnras/stw2461}, \href
  {http://adsabs.harvard.edu/abs/2017MNRAS.464.2419S} {464, 2419}

\bibitem[\protect\citeauthoryear{{Sawala} et~al.,}{{Sawala}
  et~al.}{2015}]{Sawala15}
{Sawala} T.,  et~al., 2015, \mn@doi [\mnras] {10.1093/mnras/stu2753}, \href
  {http://adsabs.harvard.edu/abs/2015MNRAS.448.2941S} {448, 2941}

\bibitem[\protect\citeauthoryear{{Sawala} et~al.,}{{Sawala}
  et~al.}{2016}]{Sawala16}
{Sawala} T.,  et~al., 2016, \mn@doi [\mnras] {10.1093/mnras/stw145}, \href
  {http://adsabs.harvard.edu/abs/2016MNRAS.457.1931S} {457, 1931}

\bibitem[\protect\citeauthoryear{{Schneider}}{{Schneider}}{2015}]{Schneider15}
{Schneider} A.,  2015, \mn@doi [\mnras] {10.1093/mnras/stv1169}, \href
  {http://adsabs.harvard.edu/abs/2015MNRAS.451.3117S} {451, 3117}

\bibitem[\protect\citeauthoryear{{Schneider}, {Smith}  \& {Reed}}{{Schneider}
  et~al.}{2013}]{Schneider13}
{Schneider} A.,  {Smith} R.~E.,   {Reed} D.,  2013, \mn@doi [\mnras]
  {10.1093/mnras/stt829}, \href
  {http://adsabs.harvard.edu/abs/2013MNRAS.433.1573S} {433, 1573}

\bibitem[\protect\citeauthoryear{{Schneider}, {Anderhalden}, {Macci{\`o}}  \&
  {Diemand}}{{Schneider} et~al.}{2014}]{Schneider14}
{Schneider} A.,  {Anderhalden} D.,  {Macci{\`o}} A.~V.,   {Diemand} J.,  2014,
  \mn@doi [\mnras] {10.1093/mnrasl/slu034}, \href
  {http://adsabs.harvard.edu/abs/2014MNRAS.441L...6S} {441, L6}

\bibitem[\protect\citeauthoryear{{Schneider}, {Trujillo-Gomez}, {Papastergis},
  {Reed}  \& {Lake}}{{Schneider} et~al.}{2017}]{Schneider17}
{Schneider} A.,  {Trujillo-Gomez} S.,  {Papastergis} E.,  {Reed} D.~S.,
  {Lake} G.,  2017, \mn@doi [\mnras] {10.1093/mnras/stx1294}, \href
  {http://adsabs.harvard.edu/abs/2017MNRAS.470.1542S} {470, 1542}

\bibitem[\protect\citeauthoryear{{Shen}, {Madau}, {Conroy}, {Governato}  \&
  {Mayer}}{{Shen} et~al.}{2014}]{Shen14}
{Shen} S.,  {Madau} P.,  {Conroy} C.,  {Governato} F.,   {Mayer} L.,  2014,
  \mn@doi [\apj] {10.1088/0004-637X/792/2/99}, \href
  {http://adsabs.harvard.edu/abs/2014ApJ...792...99S} {792, 99}

\bibitem[\protect\citeauthoryear{{Somerville}}{{Somerville}}{2002}]{Somerville02}
{Somerville} R.~S.,  2002, \mn@doi [\apjl] {10.1086/341444}, \href
  {http://adsabs.harvard.edu/abs/2002ApJ...572L..23S} {572, L23}

\bibitem[\protect\citeauthoryear{{Springel} et~al.,}{{Springel}
  et~al.}{2005}]{Springel05}
{Springel} V.,  et~al., 2005, \mn@doi [\nat] {10.1038/nature03597}, \href
  {http://adsabs.harvard.edu/abs/2005Natur.435..629S} {435, 629}

\bibitem[\protect\citeauthoryear{{Stilp}, {Dalcanton}, {Warren}, {Skillman},
  {Ott}  \& {Koribalski}}{{Stilp} et~al.}{2013}]{Stilp13}
{Stilp} A.~M.,  {Dalcanton} J.~J.,  {Warren} S.~R.,  {Skillman} E.,  {Ott} J.,
   {Koribalski} B.,  2013, \mn@doi [\apj] {10.1088/0004-637X/765/2/136}, \href
  {http://adsabs.harvard.edu/abs/2013ApJ...765..136S} {765, 136}

\bibitem[\protect\citeauthoryear{{Swaters}, {Sancisi}, {van Albada}  \& {van
  der Hulst}}{{Swaters} et~al.}{2009}]{Swaters09}
{Swaters} R.~A.,  {Sancisi} R.,  {van Albada} T.~S.,   {van der Hulst} J.~M.,
  2009, \mn@doi [\aap] {10.1051/0004-6361:200810516}, \href
  {http://adsabs.harvard.edu/abs/2009A%26A...493..871S} {493, 871}

\bibitem[\protect\citeauthoryear{{Teyssier}, {Pontzen}, {Dubois}  \&
  {Read}}{{Teyssier} et~al.}{2013}]{Teyssier13}
{Teyssier} R.,  {Pontzen} A.,  {Dubois} Y.,   {Read} J.~I.,  2013, \mn@doi
  [\mnras] {10.1093/mnras/sts563}, \href
  {http://adsabs.harvard.edu/abs/2013MNRAS.429.3068T} {429, 3068}

\bibitem[\protect\citeauthoryear{{Tikhonov} \& {Klypin}}{{Tikhonov} \&
  {Klypin}}{2009}]{Tikhonov09}
{Tikhonov} A.~V.,  {Klypin} A.,  2009, \mn@doi [\mnras]
  {10.1111/j.1365-2966.2009.14686.x}, \href
  {http://adsabs.harvard.edu/abs/2009MNRAS.395.1915T} {395, 1915}

\bibitem[\protect\citeauthoryear{{Tollet} et~al.,}{{Tollet}
  et~al.}{2016}]{nihaoIV}
{Tollet} E.,  et~al., 2016, \mn@doi [\mnras] {10.1093/mnras/stv2856}, \href
  {http://adsabs.harvard.edu/abs/2016MNRAS.456.3542T} {456, 3542}

\bibitem[\protect\citeauthoryear{{Trujillo-Gomez}, {Klypin}, {Primack}  \&
  {Romanowsky}}{{Trujillo-Gomez} et~al.}{2011}]{Trujillo-Gomez11}
{Trujillo-Gomez} S.,  {Klypin} A.,  {Primack} J.,   {Romanowsky} A.~J.,  2011,
  \mn@doi [\apj] {10.1088/0004-637X/742/1/16}, \href
  {http://adsabs.harvard.edu/abs/2011ApJ...742...16T} {742, 16}

\bibitem[\protect\citeauthoryear{{Trujillo-Gomez}, {Klypin}, {Col{\'{\i}}n},
  {Ceverino}, {Arraki}  \& {Primack}}{{Trujillo-Gomez}
  et~al.}{2015}]{Trujillo-Gomez15}
{Trujillo-Gomez} S.,  {Klypin} A.,  {Col{\'{\i}}n} P.,  {Ceverino} D.,
  {Arraki} K.~S.,   {Primack} J.,  2015, \mn@doi [\mnras]
  {10.1093/mnras/stu2037}, \href
  {http://adsabs.harvard.edu/abs/2015MNRAS.446.1140T} {446, 1140}

\bibitem[\protect\citeauthoryear{{Wang}, {Dutton}, {Stinson}, {Macci{\`o}},
  {Penzo}, {Kang}, {Keller}  \& {Wadsley}}{{Wang} et~al.}{2015}]{nihaoI}
{Wang} L.,  {Dutton} A.~A.,  {Stinson} G.~S.,  {Macci{\`o}} A.~V.,  {Penzo} C.,
   {Kang} X.,  {Keller} B.~W.,   {Wadsley} J.,  2015, \mn@doi [\mnras]
  {10.1093/mnras/stv1937}, \href
  {http://adsabs.harvard.edu/abs/2015MNRAS.454...83W} {454, 83}

\bibitem[\protect\citeauthoryear{{Warren} et~al.,}{{Warren}
  et~al.}{2012}]{Warren12}
{Warren} S.~R.,  et~al., 2012, \mn@doi [\apj] {10.1088/0004-637X/757/1/84},
  \href {http://adsabs.harvard.edu/abs/2012ApJ...757...84W} {757, 84}

\bibitem[\protect\citeauthoryear{{Wheeler}, {O{\~n}orbe}, {Bullock},
  {Boylan-Kolchin}, {Elbert}, {Garrison-Kimmel}, {Hopkins}  \& {Kere{\v
  s}}}{{Wheeler} et~al.}{2015}]{Wheeler15}
{Wheeler} C.,  {O{\~n}orbe} J.,  {Bullock} J.~S.,  {Boylan-Kolchin} M.,
  {Elbert} O.~D.,  {Garrison-Kimmel} S.,  {Hopkins} P.~F.,   {Kere{\v s}} D.,
  2015, \mn@doi [\mnras] {10.1093/mnras/stv1691}, \href
  {http://adsabs.harvard.edu/abs/2015MNRAS.453.1305W} {453, 1305}

\bibitem[\protect\citeauthoryear{{Zavala}, {Jing}, {Faltenbacher}, {Yepes},
  {Hoffman}, {Gottl{\"o}ber}  \& {Catinella}}{{Zavala} et~al.}{2009}]{Zavala09}
{Zavala} J.,  {Jing} Y.~P.,  {Faltenbacher} A.,  {Yepes} G.,  {Hoffman} Y.,
  {Gottl{\"o}ber} S.,   {Catinella} B.,  2009, \mn@doi [\apj]
  {10.1088/0004-637X/700/2/1779}, \href
  {http://adsabs.harvard.edu/abs/2009ApJ...700.1779Z} {700, 1779}

\bibitem[\protect\citeauthoryear{{Zolotov} et~al.,}{{Zolotov}
  et~al.}{2012}]{Zolotov12}
{Zolotov} A.,  et~al., 2012, \mn@doi [\apj] {10.1088/0004-637X/761/1/71}, \href
  {http://adsabs.harvard.edu/abs/2012ApJ...761...71Z} {761, 71}

\bibitem[\protect\citeauthoryear{{Zwaan}, {Meyer}  \& {Staveley-Smith}}{{Zwaan}
  et~al.}{2010}]{Zwaan10}
{Zwaan} M.~A.,  {Meyer} M.~J.,   {Staveley-Smith} L.,  2010, \mn@doi [\mnras]
  {10.1111/j.1365-2966.2009.16188.x}, \href
  {http://adsabs.harvard.edu/abs/2010MNRAS.403.1969Z} {403, 1969}

\makeatother
\end{thebibliography}


\appendix

\section{Inclination uncertainties}
\label{sec:appendix1}

\begin{figure}
 \includegraphics[width=0.48\textwidth]{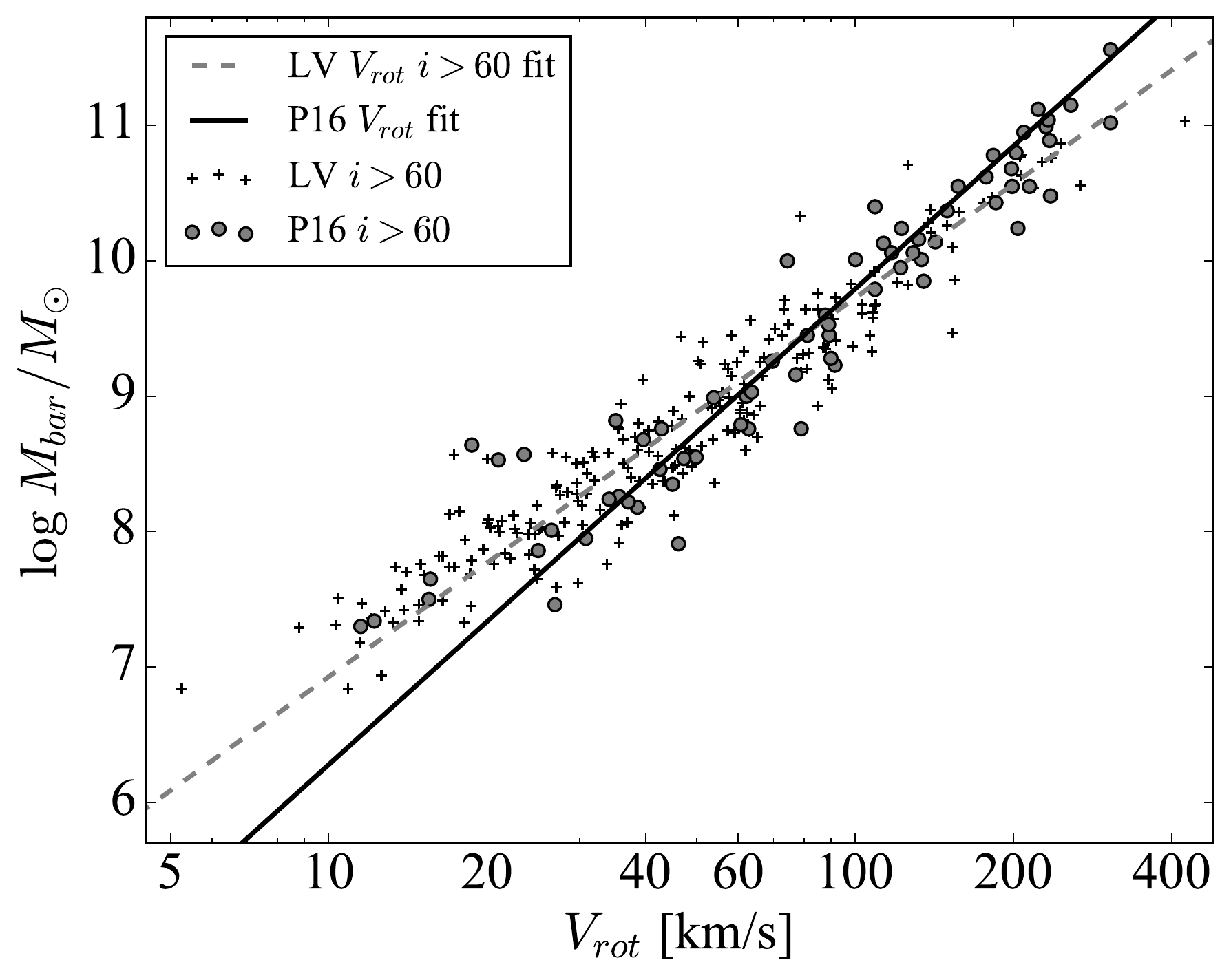} 
 \caption{BTF relation of the high-inclination objects from the LV and P16 samples. The solid circles show the P16 galaxies, while the crosses represent the LV galaxies. The solid line is the fit to the full P16 sample from Eq. \ref{eq:VrotP16}, while the dashed line is a linear fit to the high-inclination LV objects.} 
 \label{LV_and_resolved_inc}
\end{figure}

Figure~\ref{LV_and_resolved_inc} shows the BTF relations of the high-inclination ($i > 60$) LV and P16 subsamples. The inclination uncertainty in these objects is minimised compared to the full samples. Both the high-inclination LV and P16 dwarf galaxies have significantly lower rotation velocities than the full samples. As discussed by \citet{PapastergisShankar16}, this arises because the stellar discs of dwarf galaxies are thicker than those of bright spirals. For low-inclination dwarfs the assumption of infinitely thin discs can lead to underestimated inclinations, resulting in an overestimate of the deprojected rotation velocity.

To avoid this systematic effect in our analysis we used the line-of-sight rotation velocity of the LV galaxies. In addition, as discussed in Section \ref{sec:Vlos}, removing the inclination errors in the P16 sample would reduce the average $\Vout$ at a fixed $\Mbar$, allowing for lower mass haloes to be fit to the same galaxies. This would have the overall effect of a systematic shift in the observed $\Vmax$ VF towards smaller circular velocities, making the disagreement with CDM predictions even worse. 

\section{Non-parametric description of the BTFR data}
\label{sec:appendix1a}

\begin{figure}
 \includegraphics[width=0.48\textwidth]{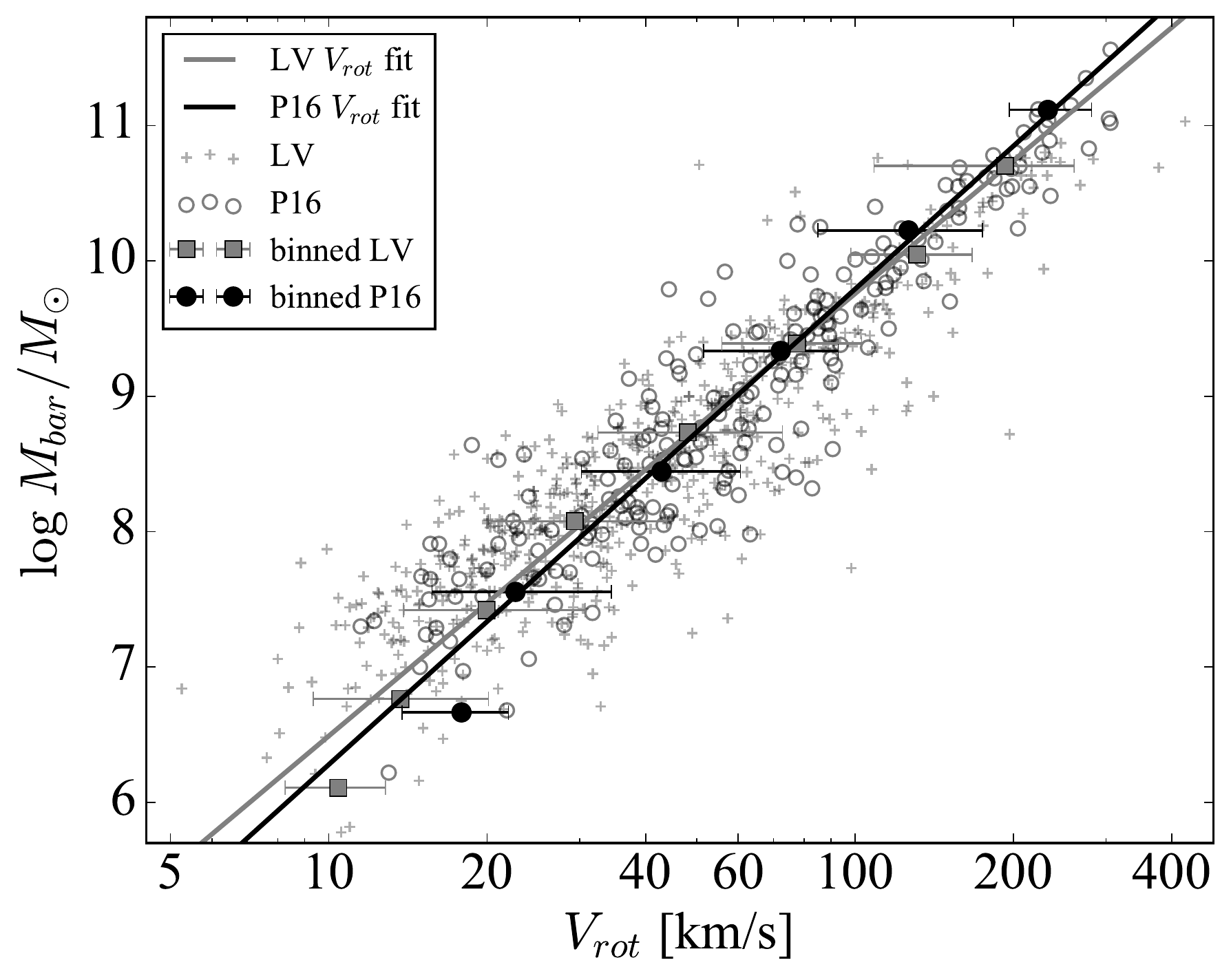} 
 \caption{Non-parametric description of the BTF relations of the Local Volume and P16 data using the deprojected \HI line-width. The symbols with error bars represent the sample mean and standard deviation in uniform $\log \Mbar$ bins. The lines reproduce the fits shown in Figure \ref{LV_and_resolved}. The power-law fits used in our analysis are an excellent representation of the non-parametric binned data except in the lowest mass bins where data is scarce.} 
\label{LV_and_resolved_binned}
\end{figure}

Figure \ref{LV_and_resolved_binned} shows a comparison of the linear regressions used to describe the BTF data in Section \ref{sec:Vlos} with the distribution of the deprojected line-widths of the LV and P16 data in uniform $\log \Mbar$ bins. The power-law fits used in our analysis appropriately capture the relation between baryonic mass and line-of-sight velocity in both samples and no extra degrees of freedom are necessary in the fits.

\section{Sensitivity of the results to sample selection}
\label{sec:appendix2}

\begin{figure*}
 \includegraphics[width=0.49\textwidth]{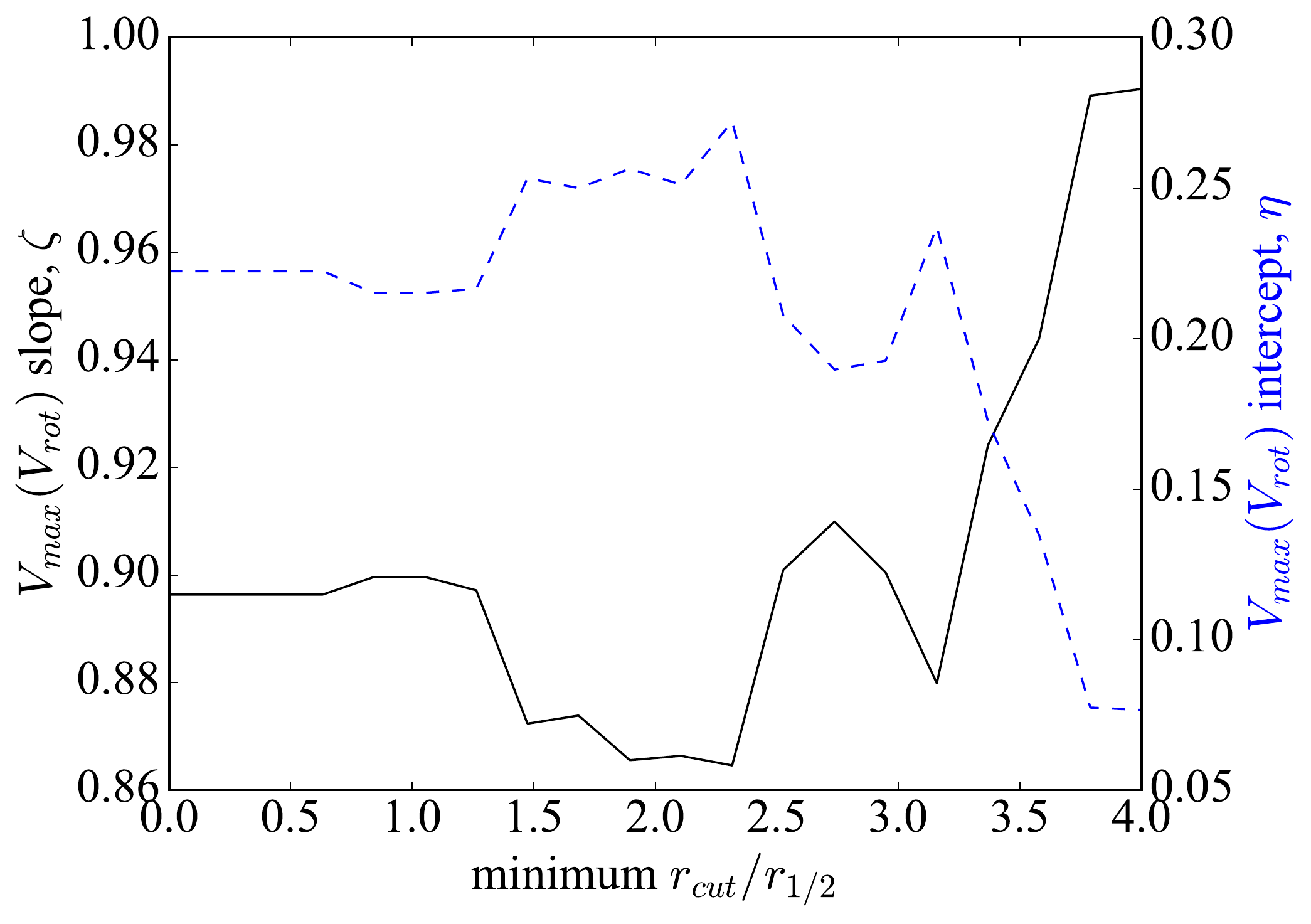} 
 \includegraphics[width=0.44\textwidth]{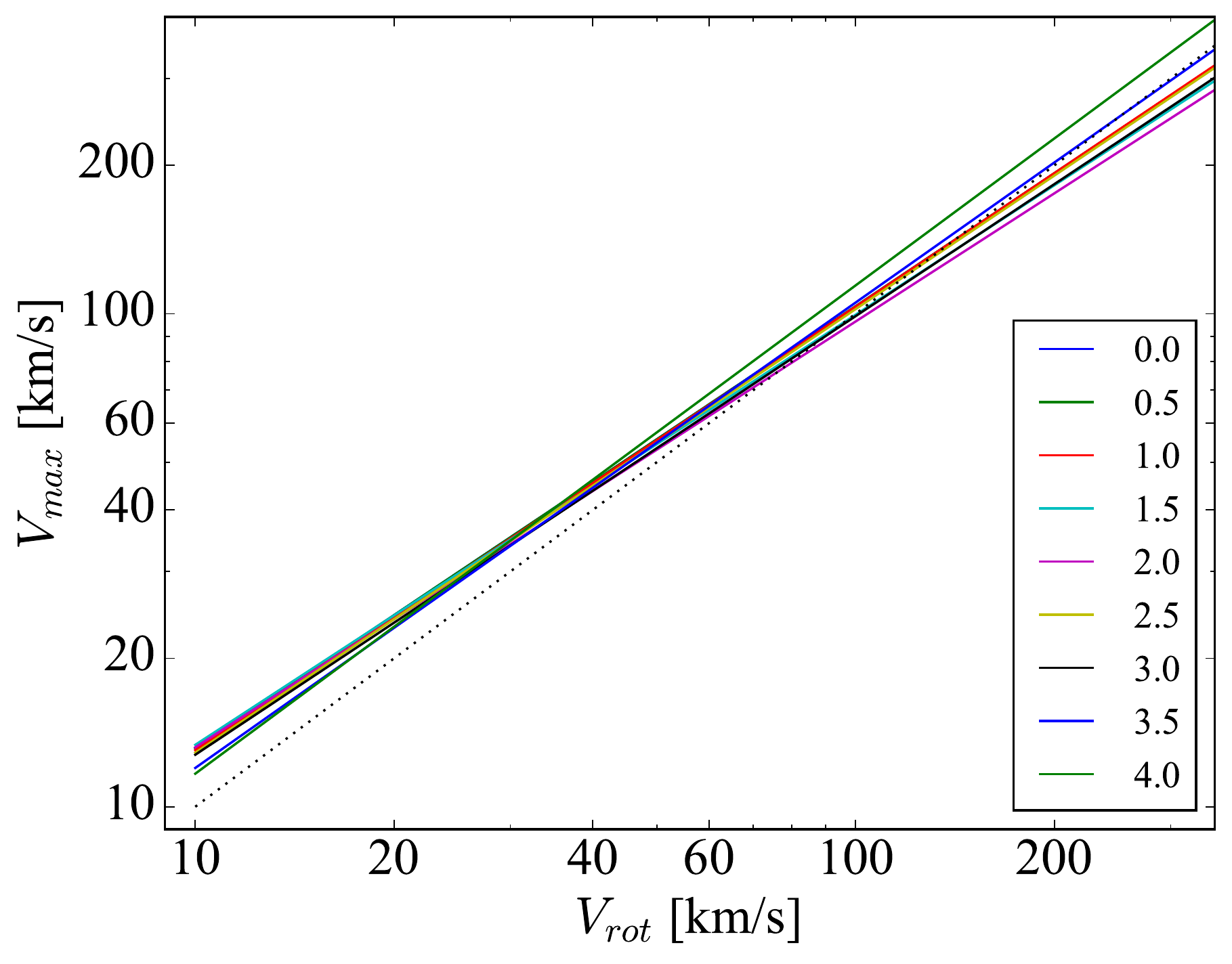}
 \caption{Effect of varying the minimum $r_{out}/r_{1/2}$ of the P16 sample on the linear fit to Eq. \ref{eq:vmax_vrot}. Left: the left axis and solid line correspond to the slope, $\zeta$, and the right axis and dashed line show the intercept, $\eta$. Right: comparison of the models. The value of $r_{\rm out}/r_{1/2}$ for each model is indicated in the legend.  The dotted line shows $\Vmax = \Vrot$. Only cuts which preserve more than 60 data points are shown. The variation in the fit parameters is relatively small, with the slope changes usually compensated by changes in the intercept to produce a robust fit at low velocities.} 
 \label{vrot_vnfw_rec1b}
\end{figure*}

Figure \ref{vrot_vnfw_rec1b} shows the sensitivity of the linear fit from  Eq. \ref{eq:vmax_vrot} to changes in the minimum value of the relative kinematic  extent, $r_{\rm out}/r_{1/2}$, allowed in the P16 sample selection. The figure shows the slope and intercept of the fit obtained from binning the error-weighted $\Vmax$ values as was done in Figure \ref{vrot_vnfw}. 

The observed variation in the fit parameters is relatively small and the largest deviations in the slope are compensated by changes in the intercept which result in a fit that is largely insensitive to the data selection.

\begin{figure}
 \includegraphics[width=0.48\textwidth]{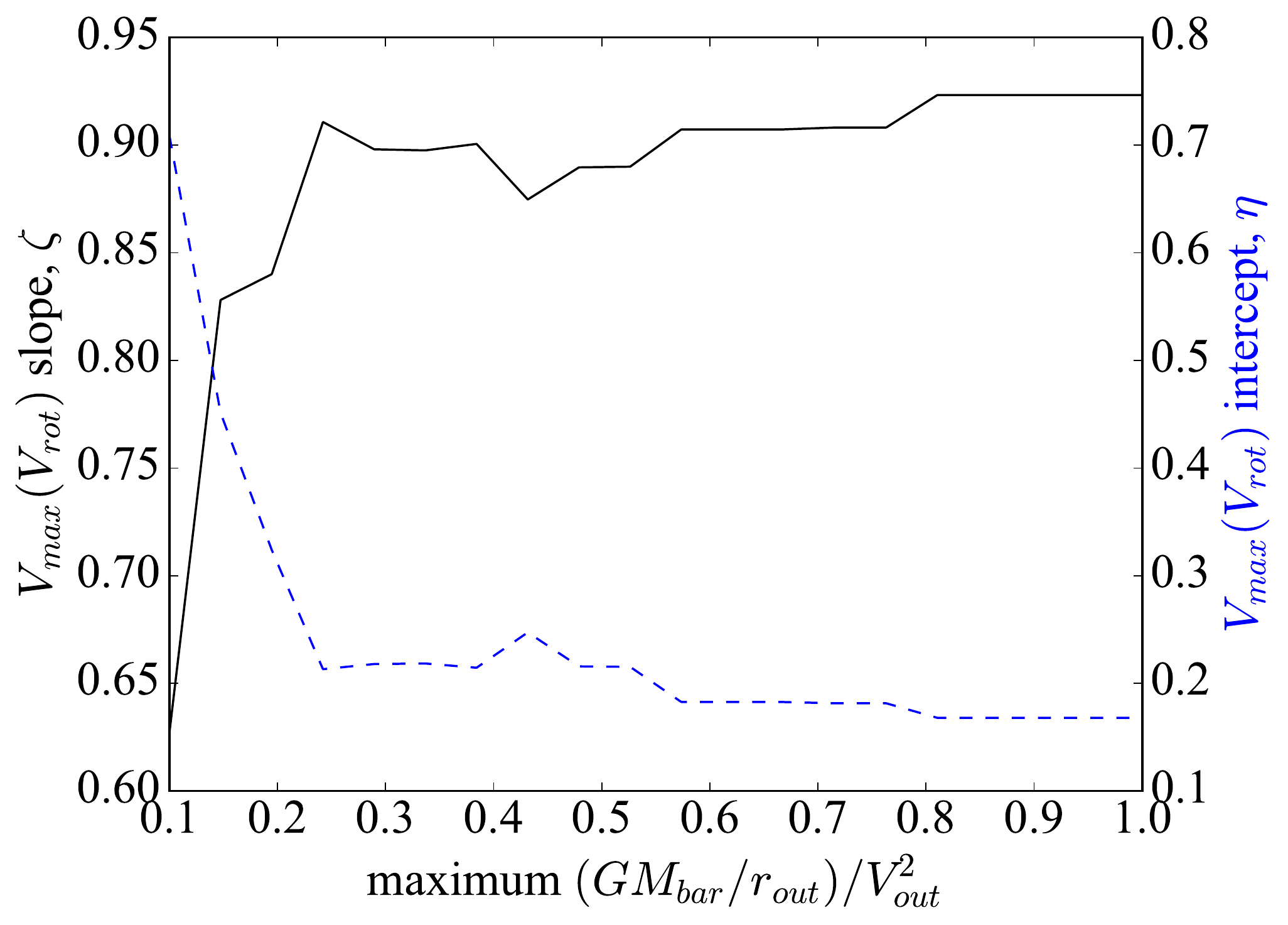} 
 \caption{Effect of varying the maximum allowed contribution from the baryonic mass to the circular velocity. The left axis and solid line correspond to the slope, $\zeta$, and the right axis and dashed line show the intercept, $\eta$. Only cuts which preserve more than 60 data points are shown. } 
 \label{vrot_vnfw_rec2b}
\end{figure}

To check the robustness of the second sample cut, we plot in Figure \ref{vrot_vnfw_rec2b} the variation of the same parameters when the maximum allowed contribution from baryons to the circular velocity, $(G\Mbar/r_{\rm out})/\vout^2$, is varied.

Although the slope and intercept vary considerably for values $(G\Mbar/r_{\rm out})/\Vout^2 < 0.2$, the intercept increases while the slope decreases to produce a relatively small net change in the predictions of the model.

\section{Comparison with other parameterisations of cored halo profiles}
\label{sec:appendix3}

\begin{figure}
 \includegraphics[width=0.65\textwidth]{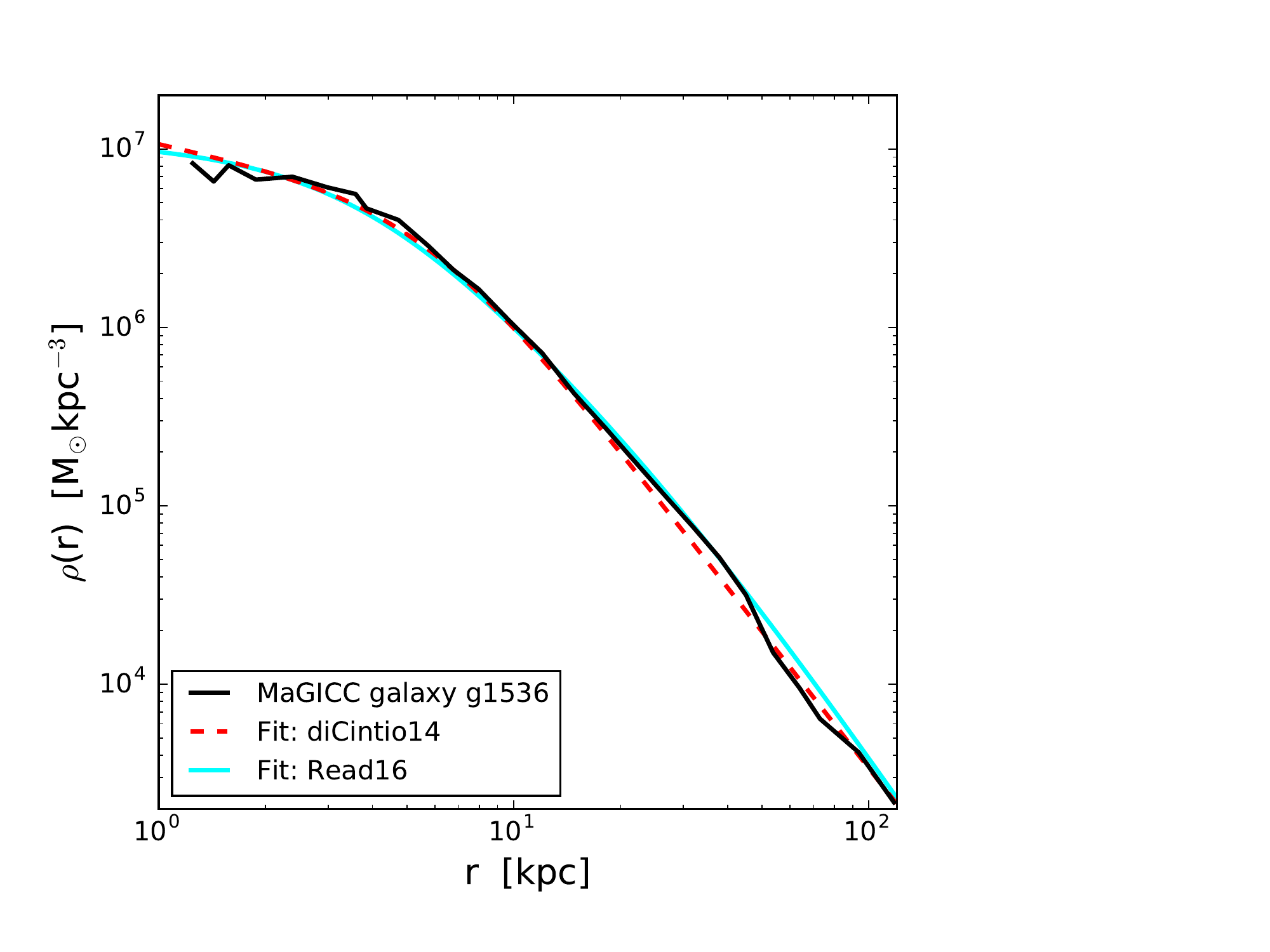} 
 \caption{Comparison of the \citet{DiCintio14} and \citet{Read16} cored DM halo profile parameterisations. The solid black line is the density profile of the hydrodynamical galaxy simulation g1536 from \citet{DiCintio14}. The dashed and solid curves are the DC14 and coreNFW fits respectively.} 
 \label{DC14vsR16}
\end{figure}

The \citet[][hereon DC14]{DiCintio14} parameterisation of the DM density profile of galaxies including feedback-induced cores has been extensively used in the literature. In this work we used instead the fit by \citet[][the ``coreNFW'' profile]{Read16} to evaluate the effect of DM cores on the observed $\vmax$ velocity function. Figure \ref{DC14vsR16} shows a comparison of the two fits applied to a simulation from the MAGICC suite \citep{DiCintio14}. Since the DC14 fit depends critically on the ratio $\Mstar/\Mvir$, we used the values reported for the simulation by \citet{DiCintio14}. For the coreNFW fit we took the half-light radius of the same simulation from \citet{Brook12}, assumed a flat core with $n=1$, and varied the concentration to obtain a good fit. The figure shows that the coreNFW fit is also a good description of the DM core formed in the simulation. 

This example further confirms the validity of the coreNFW profile for accurately capturing the extent of feedback-induced cores in hydrodynamic simulations. Our results are therefore independent of our choice of cored profile model.  

\section{How accurately does the linewidth-derived $\vrot$ trace $\vout$?}

\begin{figure}
 \includegraphics[width=0.48\textwidth]{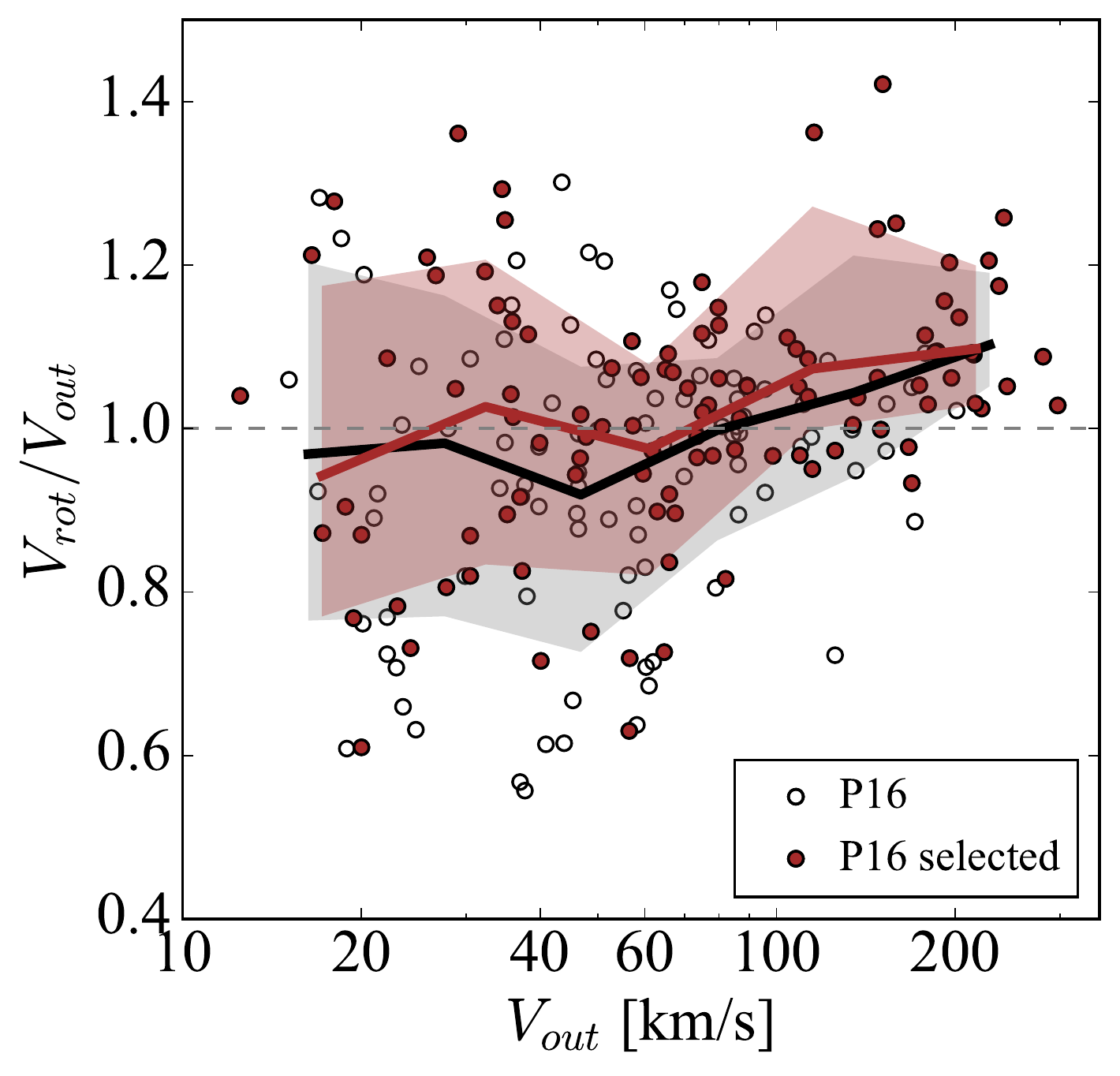} 
 \caption{Ratio between the deprojected line-of-sight velocity $\vrot$ and the outermost rotation measurement $\vout$ for the galaxies in the P16 sample. The empty symbols show the full sample while filled symbols show the selected subsample with $r_{\rm out} > 3r_{1/2}$. The solid lines and shading represent the means and r.m.s. scatter in uniform $\log \vout$ bins.} 
 \label{V50_Vout}
\end{figure}

Figure \ref{V50_Vout} shows the ratio between $\vrot$ and $\vout$ for the P16 observations. The average linewidth-derived rotation velocity traces the outermost rotation velocity of the galaxies to within $\sim 10$ per cent. 

\bsp	
\label{lastpage}
\end{document}